\documentclass[prd,10pt,aps,twocolumn,nofootinbib,floatfix]{revtex4-1}
\usepackage{amssymb}
\usepackage{mathtools}
\usepackage{amsmath}
\usepackage{slashed}
\usepackage{epsfig}

\newcommand{\beqn}{\begin{eqnarray}}
\newcommand{\eeqn}{\end{eqnarray}}
\newcommand{\beqs}{\begin{subequations}}
\newcommand{\eeqs}{\end{subequations}\\[-2mm]\noindent}
\newcommand{\eq}[1]{(\ref{#1})}

\newcommand{\ph}{{\mathrm {ph}}}
\newcommand{\mon}{{\mathrm {mon}}}
\newcommand{\cL}{{\cal L}}
\newcommand{\cZ}{{\cal Z}}
\newcommand{\cO}{{\cal O}}
\newcommand{\cH}{{\cal H}}
\newcommand{\cD}{{\cal D}}
\newcommand{\sumint}[1]{\sum_{#1}\hskip -5mm\int \hskip 2mm}

\newcommand{\Z}{{\mathbb Z}}
\newcommand{\R}{{\mathbb R}}

\newcommand{\bs}{\boldsymbol}
\newcommand{\bx}{\boldsymbol {x}}

\usepackage{xcolor}
\usepackage{ulem}
\definecolor{purple}{rgb}{0.8,0,0.6}

\begin{document}

\title{Inhomogeneous confining-deconfining phases in rotating plasmas}

\author{M. N. Chernodub}
\affiliation{Institut Denis Poisson UMR 7013, Universit\'e de Tours, 37200, Tours, France}
\affiliation{Pacific Quantum Center, Far Eastern Federal University, Sukhanova 8, 690950, Vladivostok, Russia}

\begin{abstract}
We discuss the effects of rotation on confining properties of gauge theories focusing on compact electrodynamics in two spatial dimensions as an analytically tractable model. We show that at finite temperature, the rotation leads to a deconfining transition starting from a certain distance from the rotation axis. A uniformly rotating confining system possesses, in addition to the usual confinement and deconfinement phases, a mixed inhomogeneous phase which hosts spatially separated confinement and deconfinement regions. The phase diagram thus has two different deconfining temperatures. The first deconfining temperature can be made arbitrarily low by sufficiently rapid rotation while the second deconfining temperature is largely unaffected by the rotation. Implications of our results for the phase diagram of QCD are presented. We point out that uniformly rotating quark-gluon plasma should therefore experience an inverse hadronization effect when the hadronization starts from the core of the rotating plasma rather than from its boundary. 
\end{abstract}

\date{\today}

\maketitle

\section{Introduction}

Noncentral relativistic heavy-ion collisions create a highly vortical fluid of quark-gluon plasma. According to the experimental results of the RHIC collaboration~\cite{STAR:2017ckg}, this fluid possesses the angular momentum of the order of $10^{3} \hbar$ and carries the vorticity in the range of $\omega \approx (9 \pm 1)\times 10^{21} \, \mathrm{s}^{-1} \sim 0.03\, \mathrm{fm}^{-1} c$. Theoretically, the vorticity is expected to reach even higher values, $\omega \sim 0.1\, \mathrm{fm}^{-1} c \sim 20\,\mathrm{MeV}$, at certain impact parameters~\cite{Deng:2016gyh,Jiang:2016woz}.\footnote{Hereafter, we work in units $\hbar = c = 1$.} Therefore, the vorticity can affect substantially the thermodynamic properties and the phase structure of the quark-gluon plasma. An overview of hydrodynamic and transport-based models of vortical effects in quark-gluon plasma and the relevant experimental results can be found in Refs.~\cite{Becattini:2020ngo,Huang:2020dtn}.

Theoretically, the effects of vorticity on the phase diagram are usually studied in the approximation of a uniform rotation which assumes that the quark-gluon plasma rotates as a solid body. There is a consensus that the uniform rotation decreases the temperature of the chiral phase transition~\cite{Chen:2015hfc,Jiang:2016wvv,Chernodub:2016kxh,Chernodub:2017ref,Wang:2018sur,Zhang:2020hha} because the vorticity tends to align the spins of quarks and anti-quarks and suppresses the scalar pairing thus diminishing the scalar fermionic condensate~\cite{Jiang:2016wvv}. Therefore, the effect of rotation in vortical plasmas restores the chiral symmetry at lower critical temperatures as compared to the quark-gluon plasmas with zero vorticity.

Besides the chiral symmetry breaking, QCD possesses the color confinement phenomenon in the low-temperature phase. Recent numerical lattice simulations indicate that the bulk critical temperature of the deconfining phase transition can rise with the increase of the vorticity~\cite{Braguta:2020eis}. A holographic approach gives the opposite result implying that the temperature of the deconfinement transition decreases in the vortical gluonic fluid~\cite{Chen:2020ath}. In our paper, we propose the third scenario, in which the rotation splits the single deconfining transition into the two deconfining transitions, between the pure confinement phase, a new mixed inhomogeneous phase in which the confinement and deconfinement phases are spatially separated, and the pure deconfinement phase.

To illustrate our proposal, we consider in detail the rotation in compact electrodynamics (called also compact QED or cQED for shortness) which is a toy model that possesses the confinement property and, at the same time, can be treated analytically. This (2+1) dimensional theory enjoys the ``compact'' U(1) gauge symmetry, cU(1) and it has no matter fields. The compact Abelian gauge theory in two space dimensions is an effective toy model which shares several non-perturbative features with QCD, notably the charge confinement and the mass-gap generation~\cite{Polyakov:1976fu}. Both cQED and QCD possess instanton-like objects in their vacua, and both of them experience a deconfining phase at high enough temperatures. Therefore, we expect that the effects of rotation on QCD and cQED diagrams should share similar features.

The structure of this paper is as follows. In Section~\ref{sec:rotation:scalar}, we discuss the implementation of the uniform rotation in the imaginary-time formalism in the Euclidean-space time, which is suitable for the analytical treatment of thermodynamic ensembles in thermal equilibrium. We show that the uniform rotation may be taken into account as a simple shift of the Matsubara frequencies in the complex plane. We also discuss the analytic rotation of the real-valued angular frequency to a pure imaginary domain, $\Omega \to \Omega_I = - i \Omega$, suitable for numerical lattice simulations~\cite{Yamamoto:2013zwa,Braguta:2020eis}.  In Section~\ref{sec:cU(1):short:review}, we briefly overview the non-perturbative features of compact electrodynamics in two spatial dimensions.  To elucidate the effect of rotation on confining properties of this model, we study the Green's functions in the rotating disk-shaped domain in Section~\ref{sec:greens}. These results are used in Section~\ref{sec:confinement} to find the effect of the rotation on the confining properties of the compact Abelian model. We briefly discuss the implications of our results for QCD at the end of the article. 

\section{Rotation and Wick rotation}
\label{sec:rotation:scalar}

The rotating quantum systems were originally considered by Vilenkin long time ago~\cite{Vilenkin:1980zv}. In this section, we discuss in very detail the spatial rotation in the imaginary time formalism for the simplest case of a scalar field theory. We also briefly touch the case of fermions. Our main aim is to explicitly demonstrate that the physical rotation may be incorporated as a simple shift of the Matsubara frequency in the Wick-rotated space-time. We also discuss rotation with purely imaginary frequencies $\Omega = i \Omega_I$, viewed as an analytical continuation in the plane of complex $\Omega$. We will use these results in the subsequent discussions of the confining properties of rotating compact electrodynamics.

\subsection{Thermodynamics of nonrotating scalar fields}

\subsubsection{Zero temperature}

Consider a free theory of a massless real-valued field $\phi$:
\beqn
\cL = \frac{1}{2} \partial_\mu \phi \partial^\mu \phi,
\label{eq:cL}
\eeqn
with the partition function
\beqn
\cZ = \int D \phi \, \exp\left\{ - i \int d^4 x \cL \right\} = \frac{1}{\sqrt{\mathrm{Det} \left( \Box  \right)}},
\label{eq:Z:phi}
\eeqn
and the classical equation of motion:
\beqn
\Box \, \phi(t,{\bs x}) \equiv \left(\frac{\partial^2}{\partial t} - \Delta_{\bs x} \right) \phi(t,{\bs x}) = 0,
\label{eq:class}
\eeqn
where $\Delta_{\bs x} \equiv {\bs \nabla}^2$ is the Laplacian. We traditionally ignored an irrelevant constant factor in the right-hand side of Eq.~\eq{eq:Z:phi}. In this section, we will mostly work in $3+1$ dimensions. 

The determinant in Eq.~\eq{eq:Z:phi} should be taken for the d'Alembertian operator $\Box$ defined in the appropriate spacetime. Since in our work we focus on the rotation problem in a field theory, we restrict out attention to the volume inside a cylinder of radius $R$. For simplicity, we consider the Dirichlet condition imposed on the field at the cylindrical boundary of the system:
\beqn
\phi(t,\rho,\varphi,z) {\biggl|}_{\rho = R} = 0,
\label{eq:phi:Dirichlet}
\eeqn
with the coordinate $z$ directed along the cylinder's axis, $x = \rho \cos \varphi$ and $y = \rho \sin \varphi$.

The eigenmodes are defined by Eq.~\eq{eq:class}. It is convenient to rewrite this equation in the cylindrical coordinates:
\beqn
\left(\frac{\partial^2}{\partial t} - \frac{1}{\rho} \frac{\partial}{ \partial \rho} \rho \frac{\partial}{ \partial \rho}  
- \frac{1}{\rho^2} \frac{\partial^2}{\partial \varphi^2} - \frac{\partial^2}{\partial z^2} \right) \phi_J = 0,
\label{eq:phi:eq}
\eeqn
where we introduced the cumulative index 
\beqn
J = (m, l, k_z),
\eeqn
for angular ($m \in \Z$), radial ($l \geqslant 1$), and longitudinal ($k_x \in \R$) quantum numbers that characterize the eigenfunctions:
\beqs
\beqn
\phi_J(t,\rho,\varphi,z) & = & e^{- i \omega_J t} \phi_J(\rho,\varphi,z), \\
\phi_J(\rho,\varphi,z) & = & \frac{e^{i k_z z + i m \varphi}}{\sqrt{2} \pi |J_{m+1}(\kappa_{ml})|} J_m \biggl(\kappa_{ml} \frac{\rho}{R}\biggr).
\label{eq:phi:J}
\eeqn
\label{eq:Phi:J}
\eeqs
The spatial wavefunction~\eq{eq:phi:J} is the eigenmode of the Laplacian:
\beqn
- \Delta_{\bs x} \phi_J({\bs x}) = \omega_J^2 \phi_J({\bs x}).
\label{eq:Delta:phi:J}
\eeqn

For the field satisfying the Dirichlet boundary condition~\eq{eq:phi:Dirichlet}, the dimensionless quantity $\kappa_{ml}$ is the $l$-th positive root of the Bessel function $J_m$:
\beqn
J_m (\kappa_{ml}) = 0, \qquad m \in \Z, \qquad l = 1,2,\dots\,.
\label{eq:kappa:ml}
\eeqn
The eigenenergies are as follows:
\beqn
\omega_J^2 = k_z^2 + \kappa_{ml}^2/R^2.
\label{eq:omega:J}
\eeqn

The orthonormalization of the modes~\eq{eq:phi:J}
\beqn
& & \int\limits_0^R d \rho \rho \int\limits_{-\infty}^\infty d z \int\limits_0^{2\pi} d \varphi \, \phi^*_J(t,\rho,\varphi,z)  \phi_{J'}(t,\rho,\varphi,z) \\
& & \hskip 45mm = \delta_{ll'} \delta_{mm'} \delta(k_z - k'_z), \nonumber 
\eeqn
comes as a result of the identity:
\beqn
\int_0^{1} dx \, x\, J_m(\kappa_{ml} x)  J_m(\kappa_{ml'} x) = \frac{\delta_{ll'}}{2} J^2_{m+1}(\kappa_{ml}),
\eeqn
valid for any two roots of the Bessel functions~\eq{eq:kappa:ml}.

In the volume bounded by a cylinder, the integration measure over the quantum numbers is modified with respect to the integration measure in an unbounded space. The integration over the continuous spatial momentum ${\bs k}$ is replaced by the integration over the momentum along the axis of rotation $k_z$, the sum over the angular momentum $m$ about the same axis, and the radial excitation number~$l$:
\beqn
\int \frac{d^3 k}{(2 \pi)^3} \to \sumint{J} \equiv \frac{1}{\pi R^2} \sum_{m = -\infty}^\infty \sum_{l = 1}^\infty \int_{- \infty}^\infty \frac{d k_z}{2 \pi} \,.
\label{eq:phase:space:k}
\eeqn

\subsubsection{Finite temperature}

The conventional approach to study the theory in the thermodynamic equilibrium is to perform the Wick rotation from the real time to the imaginary time:
\beqn
t \to i \tau.
\label{eq:Wick}
\eeqn
It does not mean that one should literally apply the Wick transformation~\eq{eq:Wick} to the eigenfunction~\eq{eq:Phi:J}. Instead, we identify a new operator in the left-hand-side of Eq.~\eq{eq:class} with the Wick-transformed time~\eq{eq:Wick}, and define its eigensystem:
\beqn
\left(- \frac{\partial^2}{\partial \tau} - \Delta_{\bs x} \right) \phi_{n,J}(\tau,{\bs x}) = E^2_{n,J} \phi_{n,J} (\tau,{\bs x}).
\label{eq:class:T}
\eeqn

At finite temperature $T$, the direction of the imaginary time $\tau$ is compactified to a circle of the length $\beta = 1/T$ with periodic (for bosons) boundary conditions:
\beqn
\phi_{n,J}(\tau, {\bs x}) {\biggl|}_{\tau = 0} = \phi_{n,J}(\tau, {\bs x}){\biggl|}_{\tau = \frac{1}{T}}.
\label{eq:boundary:periodic}
\eeqn
The solutions of Eqs.~\eq{eq:class:T} and \eq{eq:boundary:periodic} are:
\beqn
\phi_{n,J}(\tau, {\bs x}) & = & e^{i \omega_n \tau} \phi_J({\bs x}), 
\label{eq:chi:J}
\eeqn
where the spatial function $\phi_J$ is given in Eq.~\eq{eq:phi:J} and
\beqn
\omega_n = 2 \pi n, \qquad n \in \Z,
\label{eq:omega:n}
\eeqn
is the (bosonic) Matsubara frequency labeled by the integer~$n$. According to Eqs.~\eq{eq:chi:J} and \eq{eq:phi:J}, the spectrum of Eq.~\eq{eq:class:T} is given by the eigenvalues~\eq{eq:omega:J} of the spatial Laplacian and the Matsubara frequencies~\eq{eq:omega:n}: 
\beqn
E^2_{n,J} = \omega_n^2 + \omega_J^2.
\label{eq:E:n:J}
\eeqn

The partition function, $\cZ = e^{- F/T}$, determines the free energy,
\beqn
F & = & - \frac{1}{2} T V \sum_{n \in \Z} \sum_J\hskip -5mm\int \hskip 2mm \ln \frac{E_{n,J}^2}{T^2} \nonumber \\
& \equiv & -\frac{1}{2} T V \sum_{n \in \Z} \sum_J\hskip -5mm\int \hskip 2mm \ln \frac{\omega_n^2 + \omega_J^2}{T^2},
\label{eq:F}
\eeqn
via the eigenfunctions of the operator~\eq{eq:class:T}. Here $V$ is the volume of the system, the integration measure is given in Eq.~\eq{eq:phase:space:k}, and $T^2$ is added in the denominator for the sake of the dimensional consistency.  At finite temperature, the integral over the fourth component of the momentum is replaced in Eq.~\eq{eq:F} by the sum over the Matsubara frequencies:
\beqn
\int\limits_{-\infty}^{+\infty} \frac{d k_4}{2\pi} \to T \sum_{n \in \Z}
\eeqn

In order to take the sum~\eq{eq:F}, we use the identity~\cite{ref:Kapusta},
\beqn
\sum_{n \in \Z} \ln \left[ (2\pi n - i \gamma)^2 + \theta^2 \right] & = & \theta +  \ln \left( 1 - e^{- \theta - \gamma} \right) 
\label{eq:sum:0} \\
& & + \ln \left( 1 - e^{- \theta + \gamma} \right) + C,
\nonumber
\eeqn
extended with a parameter $\gamma$ for a future use. To take the divergent sum~\eq{eq:sum:0}, we differentiate its left-hand side and evaluate the converging expression explicitly: 
\beqn
& & \frac{\partial }{\partial \theta} \sum_{n \in \Z} \ln \left[ (2\pi n - i \gamma)^2 + \theta^2 \right] \nonumber \\
& = & \sum_{n \in \Z} \frac{2 \theta}{(2\pi n - i \gamma)^2 + \theta^2} \nonumber \\
& = & \frac{1}{2} \left[  \coth \frac{\theta + \gamma}{2} + \coth \frac{\theta - \gamma}{2}\right].
\eeqn
The integration of the above expression over $\theta$ gives us Eq.~\eq{eq:sum:0} with an (infinite) constant constant $C$ which we can safely ignore in the following.

Using Eq.~\eq{eq:sum:0} with $\theta = \omega_J/T$ and $\gamma = 0$, we get for the free energy~\eq{eq:F} the following standard expression:
\beqn
\frac{F}{V} = - \frac{1}{2}  \sumint{J} \omega + \sumint{J} \ln \left( 1 - e^{-\omega_J/T} \right),
\eeqn
where the first term corresponds to the vacuum contribution and the second term is the standard thermodynamic energy. Below, we will repeat the same derivation for the rotating system of bosons.

\subsection{Rotating scalar fields after the Wick rotation}

We consider a thermal ensemble of the scalar fields rotating rigidly and uniformly with the angular frequency $\Omega$ about the $z$ axis of the same cylinder.  We will always assume that the radius of the cylinder and the rotation frequencies satisfy the causality constraint: $|\Omega R| < 1$. Without loosing generality, we assume that the system rotates counterclockwise, $\Omega > 0$. 

The cylindrical coordinates of the laboratory reference frame, $x^\mu = (t, \varphi, \rho, z)$, are related to the reference frame co-rotating with the cylinder, 
${\tilde x}^\mu = ({\tilde t}, {\tilde \varphi}, {\tilde \rho}, {\tilde z})$ as follows:
\beqn
{\tilde t} = t, \quad
{\tilde \varphi} = [\varphi - \Omega t]_{2\pi}, \quad 
{\tilde \rho} = \rho, \quad
{\tilde z} = z, \quad
\label{eq:change:nothing}
\eeqn
where $[\dots]_{2\pi}$ means modulo $2\pi$. All coordinates remain the same, except for the angle $\varphi$ which is a uniformly growing (up to $2 \pi$) linear function of time. 

The change of variables~\eq{eq:change:nothing} leads to the relation between the derivatives in these frames:
\beqn
\frac{\partial }{\partial  t} = \frac{\partial }{\partial {\tilde t}} {-} \Omega \frac{\partial }{\partial {\tilde\varphi}}, \ \ 
\frac{\partial }{\partial \varphi} = \frac{\partial }{\partial {\tilde\varphi}}, \ \ 
\frac{\partial }{\partial \rho} = \frac{\partial }{\partial {\tilde\rho}}, \ \ 
\frac{\partial }{\partial z} = \frac{\partial }{\partial {\tilde z}}.  \qquad
\label{eq:change}
\eeqn
The change of the angular coordinate affect the derivative with respect to the time coordinate. 

One should distinguish between (i) a description of a truly static (``static'', in the laboratory reference frame) system in rotating coordinates and (ii) a physically rotating system in the appropriate co-rotating coordinates. The uniform rotation is determined by the frame in which the ``matter'' (set by nonzero chemical potentials and/or nonvanishing temperature) is identified. Our ``tilted'' coordinate system is set in the co-rotated, physical reference frame. Thus, the wave-functions and the energy spectrum -- the latter will set the distribution functions of the matter -- are to be determined in the co-rotating (``tilted'') reference frame.

Substituting the relations~\eq{eq:change} into the equation of motion~\eq{eq:phi:eq}, we arrive at
\beqn
\left[ \left(\frac{\partial}{\partial {\tilde t}} {-} \Omega \frac{\partial }{\partial {\tilde\varphi}} \right)^2 - \Delta_{\tilde {\bs x}}\right] \phi_J ({\tilde t}, {\tilde{\bs x}}) = 0,
\qquad
\label{eq:phi:eq:rotation}
\eeqn
where 
\beqn
\Delta_{\tilde {\bs x}} = \frac{1}{{\tilde \rho}} \frac{\partial}{ \partial {\tilde \rho}} {\tilde \rho} \frac{\partial}{ \partial {\tilde\rho}} + \frac{1}{{\tilde\rho}^2} \frac{\partial^2}{\partial {\tilde\varphi}^2} + \frac{\partial^2}{\partial {\tilde z}^2},
\eeqn
is the spatial Laplacian in the rotating frame which is unchanged up to the trivial renaming of the coordinates.

The Wick rotation~\eq{eq:Wick} applied to the co-rotating (tilted) time coordinate leads to the modification of the differential operator in the right-hand side of Eq.~\eq{eq:phi:eq:rotation}:
\beqn
\left[- \left(\frac{\partial}{\partial {\tilde \tau}} - i \Omega \frac{\partial }{\partial {\tilde\varphi}} \right) 
- \Delta_{\tilde {\bs x}} \right] \phi_{n,J}
= {\tilde E}^2_{n,J} \phi_{n,J}.
\label{eq:class:T:rot}
\eeqn

The important but obvious point now is to notice that the periodicity of the imaginary-time direction is imposed in the co-rotating frame and not in the laboratory frame:
\beqn
\phi_{n,J}({\tilde \tau}, {\tilde {\bs x}}) & = & e^{i \omega_n {\tilde \tau}} \phi_J({\tilde {\bs x}}), 
\label{eq:chi:J:tilded}
\eeqn
The energy spectrum in the corotating frame~\eq{eq:class:T:rot} is:
\beqn
{\tilde E}^2_{n,J} = (\omega_n - i \Omega m)^2 + \omega_J^2.
\label{eq:E:n:J:rot}
\eeqn
Here we took into account the particular form of the angular dependence of the spatial wavefunction~\eq{eq:phi:J}.

The free energy in the rotating frame $\tilde F$ is defined analogously to Eq.~\eq{eq:F} but now with the shifted spectrum~\eq{eq:E:n:J:rot}. Using the equality~\eq{eq:sum:0} with $\theta = \omega_J/T$ and $\gamma = \Omega m/T$, we get for the complex bosonic field: 
\beqn
& & \frac{1}{V} {\tilde F}^{\mathrm{(b)}} = - \frac{T}{2} \sum_n \sum_J\hskip -5mm\int \hskip 2mm \ln \frac{{\tilde E}_{n,J}^2}{T^2}  = \frac{T}{2} \sumint{J} \omega 
\label{eq:F:bos:rot}\\
&  &  + \frac{T}{2} \sumint{J} \left[ \ln \left( 1 {-} e^{-(\omega_J - m \Omega)/T} \right) + \ln \left( 1 {-} e^{-(\omega_J + m \Omega)/T} \right)  \right]\!.
\nonumber
\eeqn
We find that the physical rotation in the Wick-rotated Euclidean formulation of the field theory in the thermal equilibrium is given, for a bosonic field, by a simple shift of the Matsubara frequency: $\omega_n \to \omega_{n} - i \Omega m$. In the rest of this section we will briefly rederive this relation in another, more formal way. From now on we remove the tilted marks from the co-rotating coordinates since the reference frame in which we work should be clear from the context.

In the rotating reference frame~\eq{eq:change:nothing}, the Lagrangian of the model~\eq{eq:cL} takes the following form:
\beqn
\cL = \frac{1}{2} [(\partial_t - \Omega \partial_{\varphi} )\phi]^2 - \frac{1}{2}  ({\bs\nabla}_{{\bs x}} \phi)^2.
\eeqn
The momentum conjugate to the field,
\beqn
\pi = \frac{\partial \cL}{\partial (\partial_t \phi)} = \partial_t \phi - \Omega \partial_\varphi \phi,
\eeqn
leads us to the following expression for the Hamiltonian in the rotating frame:
\beqn
\cH(\pi,\phi) = \pi \partial_t \phi - \cL = \frac{1}{2} \pi^2 + ({\bs\nabla}_{{\bs x}} \phi)^2 - \Omega \pi \partial_\varphi \phi.
\eeqn

Following the standard approach~\cite{ref:Kapusta}, we perform the Wick rotation~\eq{eq:Wick} and formulate the path integral in a discretized space-time. In the terms of the continuous variables, the path integral reads as follows:
\beqn
\cZ {=} \int \hskip -1mm D \pi \hskip -4mm \int\limits_{\mathrm{periodic}} \hskip -4mm D \phi \exp\Biggl\{  \int\limits_0^{1/T} \! d \tau \! \int \! d^3 x \Bigl( i \pi \partial_\tau \phi {-} \cH (\pi,\phi) \Bigr) \Biggr\} \quad\
\label{eq:Z:2}
\eeqn
where the expression in the exponent is
\beqn
i \pi \partial_\tau \phi {-} \cH (\pi,\phi) = - \frac{\pi^2}{2} {-} \frac{1}{2}  ({\bs\nabla}_{{\bs x}} \phi)^2 {-} i \pi \bigl( \partial_\tau {-} i \Omega \partial_\varphi \phi \bigr).\qquad
\eeqn
Substituting this expression into Eq.~\eq{eq:Z:2} and performing the integral over momentum conjugate, we arrive to the following partition function:
\beqn
\cZ {=} \hskip -3mm \int\limits_{\mathrm{periodic}} \hskip -4mm D \phi \exp\Biggl\{ \int\limits_0^{1/T} \! d \tau \! \int \! d^3 x \left[
\bigl( \partial_\tau {-} i \Omega \partial_\varphi \phi \bigr)^2 {+} ({\bs\nabla}\phi)^2\right]\Biggr\}.\qquad
\label{eq:Z:3}
\eeqn
There is no charge conjugation for the scalar fields under the exponential because the fields are real-valued quantities. We perform the Gaussian integration over the scalar fields, take into account the complex energies~\eq{eq:E:n:J:rot}, and come back to the expected expression~\eq{eq:F:bos:rot} for the real-valued free energy.

We also give, without derivation, the expression for the free energy of a uniformly rotating fermion ensemble: 
\beqn
\frac{1}{V} {\tilde F}^{\mathrm{(f)}} =  - \sumint{J} \omega & - & T \sumint{J} \left[ \ln \left( 1 {+} e^{-[\omega_J - (m + 1/2) \Omega]/T} \right) \right.
\nonumber\\
&  & +  \left. \ln \left( 1 {+} e^{-[\omega_J + (m + 1/2) \Omega]/T} \right)  \right]\!,
\label{eq:F:bos:ferm}
\eeqn
which also includes the double degeneracy factor of fermion's spin. At zero charge density (in the absence of chemical potentials), the two terms of the bosonic free energy~\eq{eq:F:bos:rot} and, separately, two terms in the fermionic free energy~\eq{eq:F:bos:ferm}, are equal to each other. The rigorous treatment of the rigidly rotating fermionic ensembles inside a cylindrical cavity may be found in Refs.~\cite{Ambrus:2014uqa,Ambrus:2015lfr}

\subsection{Imaginary angular momentum and spin-statistics theorem}

In this section, we discuss certain symmetries of the thermal theory with imaginary angular momentum in the imaginary time formalism.

We have established that a uniform rotation of a system in a thermodynamic equilibrium in Minkowski space-time may also be treated in the finite-temperature imaginary time formalism using a simple shift of the Matsubara frequencies. Generalizing our results, we write the prescription both for boson and fermions, respectively:
\beqn
\omega^{\mathrm{(b)}}_n & \to & \omega^{\mathrm{(b)}}_n - i \Omega m,  
\label{eq:omega:b}\\ 
\omega^{\mathrm{(f)}}_n & \to & \omega^{\mathrm{(f)}}_n - i \Omega \left( m + \frac{1}{2} \right),
\label{eq:omega:f}
\eeqn
where $\Omega$ is the rotation frequency, and 
\beqn
\omega^{\mathrm{(b)}}_n = 2 \pi T n, \qquad  \omega^{\mathrm{(f)}}_n = 2 \pi T \left( n + \frac{1}{2} \right),
\eeqn
are the bosonic [marked by the superscript ``(b)''] and fermionic [marked by the superscript ``(f)''] Matsubara frequencies, respectively.

Below we will treat the rotational frequency $\Omega$ as a complex variable, concentrating, in particular, at the purely imaginary quantity
\beqn
\Omega = i \Omega_I,
\label{eq:Omega:I}
\eeqn
which is suitable in first-principle lattice simulations of rotating systems~\cite{Yamamoto:2013zwa,Braguta:2020eis}.

The shift rules~\eq{eq:omega:b} and \eq{eq:omega:f} have a universal nature in a sense that they change the corresponding imaginary frequency in all instances it appears in the partition function $\cZ$ and in the thermal expectation value of any operator $\cO$. This universality allows us to establish immediately the discrete symmetry of the partition function under the shifts of the imaginary frequency~\eq{eq:Omega:I}:
\beqn
\cO(\Omega_I + 2 \pi T \ell) & = & \cO(\Omega_I), \qquad \mbox{(bosons)}
\label{eq:shift:b}\\
\cO(\Omega_I + 4 \pi T \ell) & = & \cO(\Omega_I), \qquad \mbox{(fermions)}
\label{eq:shift:f}
\eeqn
for any integer number $\ell \in \Z$. Notice that the fermion's period is twice larger than the period for bosons.

In order to illustrate the properties of the system at the imaginary angular frequency, let us consider the sums which often appear in finite-temperature calculations in bosonic and fermionic systems, respectively:
\beqn
& & S^{\mathrm{(b)}}_m (\Omega) = T \sum_{n\in\Z} \frac{1}{\left( \omega_n^{\mathrm{(b)}} - i \Omega m \right)^2 + \varepsilon_p^2} 
\label{eq:S:b}\\
& & \hskip 5mm = \frac{1}{2 \varepsilon_p}  \left[ 1 + \frac{1}{e^{(\varepsilon_p + \Omega m)/T}-1}  + \frac{1}{e^{(\varepsilon_p - \Omega m)/T}-1}  \right],\nonumber\\
& & S^{\mathrm{(f)}}_m (\Omega) = T \sum_{n\in\Z} \frac{1}{\left( \omega_n^{\mathrm{(f)}} - i \Omega \left( m + \frac{1}{2} \right) \right)^2 + \varepsilon_p^2} 
\label{eq:S:f}\\
& & = \frac{1}{2 \varepsilon_p}  \left[ 1 {-} \frac{1}{e^{\left(\varepsilon_p + \Omega (m + \frac{1}{2}) \right)/T} + 1} {-} \frac{1}{e^{\left(\varepsilon_p - \Omega (m + \frac{1}{2}) \right)/T} + 1}  \right]. \nonumber
\eeqn
Here the subscript $p$ in the dispersion relation (a particle or quasiparticle energy) $\varepsilon_p$ incorporates all quantum numbers including the angular momentum $m$. The sum over the Matsubara frequencies gives us, respectively, the bosonic and fermionic occupation numbers with the energies shifted by the angular momenta. 

The symmetries \eq{eq:shift:b} and \eq{eq:shift:f} in the momentum space are enforced by the trivial identities (with $m,\ell \in \Z$):
\beqn
e^{\frac{i \Omega_I}{T} m} {\biggl|}_{\Omega = 2 \pi T \ell} = e^{\frac{i \Omega_I}{T} \left( m + \frac{1}{2} \right)}{\biggl|}_{\Omega_I = 4 \pi T \ell} = 1.
\eeqn

The relations~\eq{eq:shift:b} and \eq{eq:shift:f} have the following consequences: the bosonic (fermionic) system with the imaginary angular momentum $\Omega_I = 2 \pi \ell$ ($\Omega_I = 4 \pi \ell$) corresponds to a nonrotating system with $\Omega = 0$:
\beqn
\cZ^{\mathrm{(b)}}(\Omega){\biggl|}_{\Omega = 2 \pi i T} & = & \cZ^{\mathrm{(b)}}(\Omega){\biggl|}_{\Omega = 0} , \qquad \\
\cZ^{\mathrm{(f)}}(\Omega){\biggl|}_{\Omega = 4 \pi i T} & = & \cZ^{\mathrm{(f)}}(\Omega){\biggl|}_{\Omega = 0} .\qquad 
\eeqn
These relations establish the equivalence between
\begin{itemize}
\item[(i)] the bosonic (fermionic) system which rotates in the imaginary space exactly once (exactly twice) as the Matsubara imaginary time makes a full circle from $\tau = 0$ to $\tau = 1/T$;
\item[(ii)] the very same system which does not rotate at all.
\end{itemize}

These equivalences have a simple interpretation in terms of the spin-statistics theorem: the bosonic wave-function transforms to itself after any number of full rotations about a fixed point while the fermionic wavefunction needs an even number of full rotations to restore its original form. Our conclusions in this subsection have a universal character: they are also valid in interacting theories because the bosonic~\eq{eq:omega:b} and fermionic~\eq{eq:omega:f} combinations appear in any loop order. 

The spin-statistics relation lead also to some consequences at the imaginary angular momentum corresponding to a half rotation. Namely, the bosonic theory at the half imaginary period, $\Omega_I = \pi T$, corresponds to an exotic non-rotating ($\Omega = 0$) theory in which the modes with odd even angular momentum ($m = 2 l$ with $l \in \Z$) correspond to bosonic modes with the Matsubara frequencies $\omega_n = 2 \pi T n$ while the odd modes ($m = 2 l + 1$) behave as fermions  (with $\omega_n = 2 \pi T (n+1/2)$, respectively) which contribute to the free energy, however, with a wrong sign. 

\begin{figure}[!thb]
\centerline{\includegraphics[scale=0.375,clip=true]{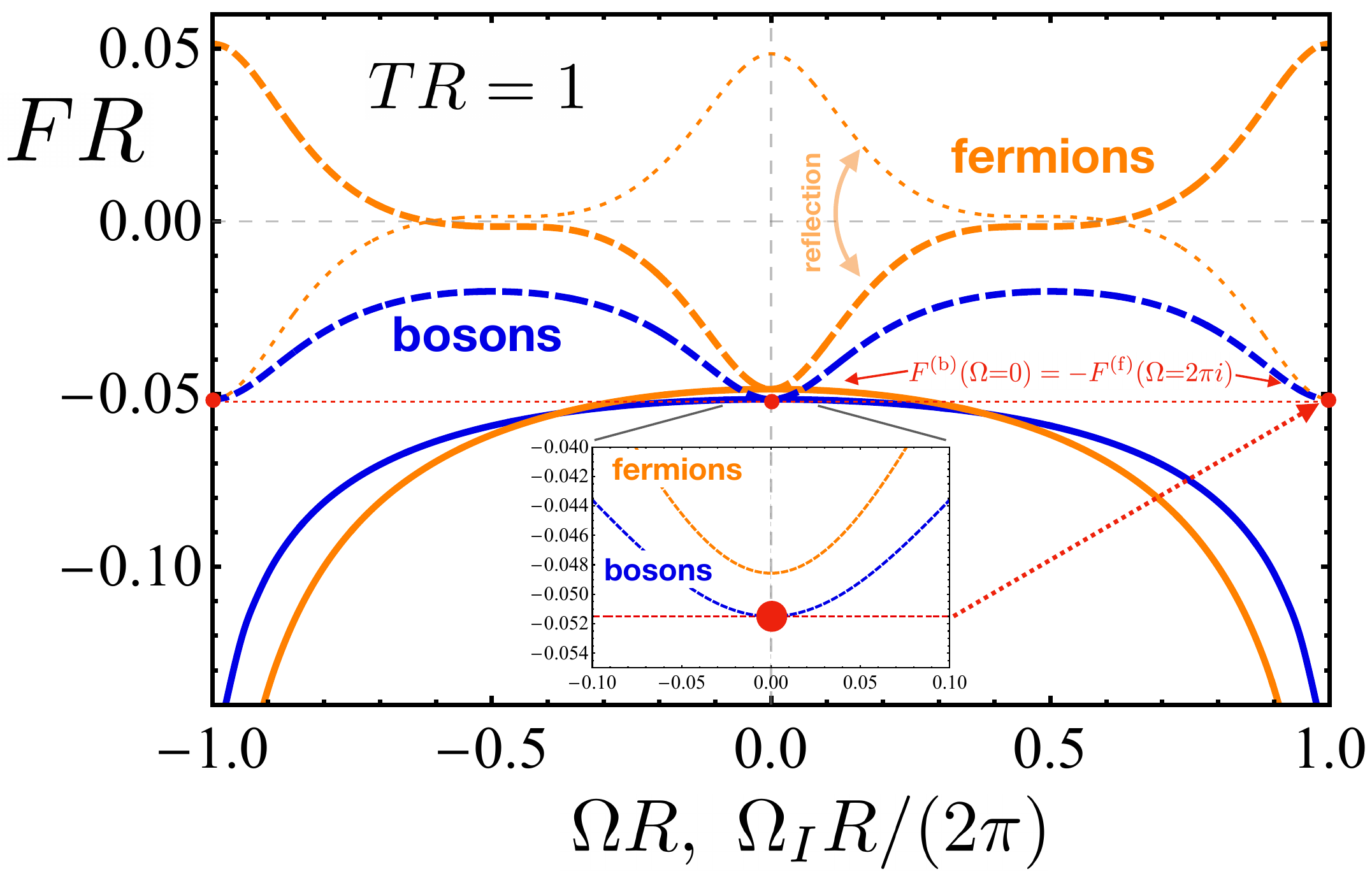}}
\caption{The free energy of bosons (blue) and fermions (orange) for real-valued ($\Omega$, the solid lines) and imaginary ($\Omega_I$, the dashed lines) angular momentum at temperature $T = R^{-1}$ in the two-dimensional disk of radius $R$. The duality~\eq{eq:F:duality} is shown via the red points at $\Omega_I = \Omega = 0$ and $\Omega_I = \pm 2\pi T$ with the same value of the free energy (shown by the horizontal dotted red line). The inset zooms in the central region and demonstrates the free energies vs. $\Omega_I$.}
\label{fig:free:energies}
\end{figure}

For fermions, the situation is more straightforward but not less exotic: at the half-period imaginary rotation in the imaginary time (corresponding to a single full rotation in the real time), $\Omega_I = 2 \pi T$, the system becomes purely bosonic with the Matsubara frequencies $\omega_n = 2 \pi T n$. However, these bosons are ghosts as they contribute to the free energy with a wrong sign.  The interpretation of this result is rather natural: a single full $2\pi$ rotation flips the sign of the fermionic wavefunction as the fields evolve the full period $1/T$ along the imaginary time. Therefore the anti-periodic boundary conditions, imposed on fermions  in the compactified imaginary time, become the periodic boundary conditions for these new bosons. For example, for the imaginary angular frequency $\Omega_I = 2 \pi T$ the system is it described by spinors obeying bosonic statistics. One thus gets:
\beqn
F^{\mathrm{(f)}}(\Omega=2\pi i) = F^{\mathrm{(gh)}}(\Omega=0) \stackrel{\mbox{\tiny{free}}}{=} - F^{\mathrm{(b)}}(\Omega=0), \quad
\label{eq:F:duality}
\eeqn
where the last equality is written for free particles. The duality~\eq{eq:F:duality} is illustrated in Fig.~\ref{fig:free:energies} for a free bosonic theory on a two-dimensional disk.

The properties of the Green's functions will be essential for the rotating plasma in compact electrodynamics. We overview the confining properties of this model in the next section.

\section{(2+1) compact electrodynamics}
\label{sec:cU(1):short:review}

At the analytical level, the confinement phenomenon is well understood in compact electrodynamics in two spatial dimensions~\cite{Polyakov:1976fu}. While this model has no dynamical matter fields, it possesses instanton-like Abelian monopoles which emerge as an inevitable result of the compactness of the U(1) gauge group. In this section, we very briefly summarize certain important features of cU(1) gauge theory concentrating on the role of the Abelian monopoles.

\subsection{Monopoles, photons, and boundaries}

\subsubsection{Lagrangian}

The compact electrodynamics is a pure U(1) gauge theory with monopoles-like singularities in the gauge field. In (3+1) spacetime dimensions, the  world trajectories of the monopoles are closed lines so that the monopoles share similarly with standard pointlike particles. This model serves is a precursor of effective approaches~\cite{Maedan:1988yi} to the confinement problem of Yang-Mills theory~\cite{Mandelstam:1974pi,tHooft:1981bkw} based the Abelian-monopole confinement mechanism which utilizes the Abelian dominance phenomenon~\cite{Kronfeld:1987vd} (for recent developments, see Ref.~\cite{Hiraguchi:2020tru}). In (2+1) dimensions, the monopoles are instanton-like objects as their trajectories are represented by a set of individual points. 

The action of the compact U(1) gauge theory,
\beqn
S [A,\varrho] = \frac{1}{4}  \int d^3 x \, F^2_{\mu\nu}
\label{eq:S:continuum}
\eeqn
is quadratic in the field strength tensor ($\mu,\nu = 1,2,3$):
\beqn
F_{\mu\nu} = F_{\mu\nu}^{\ph} + F_{\mu\nu}^{\mon}\,.
\label{eq:F:munu}
\eeqn
This tensor is the sum of the perturbative term expressed via the vector photon (gauge) field $A_\mu$,
\beqn
F_{\mu\nu}^{\ph}[A] = \partial_\mu A_\nu - \partial_\nu A_\mu\,,
\label{eq:F:ph}
\eeqn
and the nonperturbative monopole part,
\beqn
F_{\mu\nu}^{\mon}(\bx) = - g_\mon \epsilon_{\mu\nu\alpha} \partial^\alpha \int d^3 x' \, G({\bs x}, {\bs x}' ) \varrho({\bs x}')\,,
\label{eq:F:mon}
\eeqn
determined via the local monopole density:
\beqn
\varrho(\bx) = \sum_a q_a \delta^{(3)} \left(\bx - \bx_a\right)\,.
\label{eq:rho}
\eeqn 
The monopoles carry the quantized magnetic charge,
\beqn
g_\mon = \frac{2 \pi}{g}\,,
\label{eq:g:m}
\eeqn
where $g$ is the elementary electric charge. In Eq.~\eq{eq:rho}, the sum goes over the monopoles located at the positions ${\bs x}_a$ of the Euclidean spacetime with coordinates ${\bs x} = (x, y, \tau)$. We study the theory in thermal equilibrium in three-dimensional Euclidean spacetime after performing the Wick rotation, $t \to i \tau$, to the purely imaginary time. The magnetic charges $q_a \in \Z$ are written in terms of the elementary charge~\eq{eq:g:m}. We will work in the dilute gas approximation where the overlaps of the individual elementary monopoles are rare and thus $q_a = \pm 1$. Below we will take $q_a^2 = 1$.

The quantity $G({\bs x}, {\bs x}') \equiv G({\bs x}', {\bs x})$ in Eq.~\eq{eq:F:mon} is the Green's function of the three-dimensional Laplacian operator $\Delta_{\bs x} = {\bs \nabla}^2_{\bs x} \equiv \partial_\mu^2$:
\beqn
\Delta G({\bs x},{\bs x}') = - \delta({\bs x} - {\bs x}')\,.
\label{eq:Delta:G}
\eeqn
In the unbounded $\R^3$ spacetime, the Green's function,
\beqn
G({\bs x},{\bs x}') & = & G({\bs x} - {\bs x}'), 
\eeqn
is the function of a single coordinate:
\beqn
G({\bs x})  & = & \int \frac{d^3 p}{(2 \pi)^3} \frac{e^{i {\bs p} \bx}}{{\bs p}^2} = \frac{1}{4 \pi | \bx |}\,.
\label{eq:D}
\eeqn
In our paper, we will mostly work in the bounded spaces, where the Green's function becomes more involved.

In a closed space with boundaries, the action of the model~\eq{eq:S:continuum} may be written as a sum
\beqn
S [A,\varrho] & = & \frac{1}{4} \int_V d^3 x \, \bigl(F^\ph_{\mu\nu}[A] + F^\mon_{\mu\nu}[\varrho] \bigr)^2 \nonumber \\
& \equiv &
S^\ph[A] + S^\mon[\varrho]+ S^{\mathrm{surf}}[A,\varrho]\,, 
\label{eq:S:A:rho}
\eeqn
of the perturbative photon part,
\beqn
S^\ph[A] & = & \frac{1}{4} \int_V d^3 x \, \bigl(F^\ph_{\mu\nu}[A]\bigr)^2\,, 
\label{eq:S:A}
\eeqn
the non-perturbative monopole part,
\beqn
S^\mon[\varrho] & = & \frac{g^2_\mon}{2} \int_V d^3 x \, \int_V d^3 x' \, \varrho({\bs x}) G({\bs x},{\bs x}') \varrho({\bs x}'), \quad
\label{eq:S:rho}
\eeqn
and the photon-monopole surface term:
\beqn
S^{\mathrm{surf}}[A,\varrho] & = & - g_{\mon} \int_S d^2 x \, \epsilon_{\mu\nu\alpha} n^\mu({\bs x}) A^\nu({\bs x}) \nonumber \\
& & \times \, \partial^\alpha_{\bs x} \int_V d^3 x' G({\bs x},{\bs x}') \varrho({\bs x}'),
\label{eq:S:surf}
\eeqn
where $n^\mu({\bs x})$ is the local normal vector to the boundary $S = \partial V$ of the system's volume $V$. In deriving Eq.~\eq{eq:S:surf}, we used the divergence-free property of the monopole field strength: $\partial_\mu F^{\mu\nu}_\mon \equiv 0$.

\subsubsection{Boundaries}

The surface term~\eq{eq:S:surf} is an unnecessary complication of the model in a finite volume which may be removed by imposing the appropriate boundary conditions for the monopole and/or gauge fields.
In our work, we impose the Dirichlet condition for the Green's function which enters the monopole field strength~\eq{eq:F:mon}:
\beqn
G({\bs x},{\bs x}')  {\biggl|}_{{\bs x} \in S} = 0.
\label{eq:Dirichlet:G}
\eeqn
Since the derivative in Eq.~\eq{eq:S:surf} is tangential to the surface, ${\bs n}({\bs x}) \times {\bs \nabla}_{\bs x} G({\bs x},{\bs x}') = 0 $ for ${\bs x} \in \partial  S$, the Dirichlet condition~\eq{eq:Dirichlet:G} requires that the surface term~\eq{eq:S:surf} vanishes at the boundary.

Physically, the condition \eq{eq:Dirichlet:G} implies that the monopole field strength tensor~\eq{eq:F:mon} is subjected to the boundary condition:
\beqn
n^{\mu}({\bs x}) F^{\mon}_{\mu\nu}({\bs x}) {\biggl|}_{{\bs x} \in S} = 0,
\label{eq:MIT}
\eeqn
which is nothing but the MIT boundary condition for the gauge field. In three spatial dimensions, this condition would imply the vanishing of the normal electric field, ${\bs n}({\bs x}) \cdot {\bs E}({\bs x}) = 0$ as well as the tangential magnetic field, ${\bs n}({\bs x}) \times {\bs B}({\bs x}) = 0$, with ${\bs x} \in S$. The MIT boundary conditions correspond to a boundary made of a perfect magnetic conductor,  which is electromagnetically dual to a perfectly electrically conducting material.

\subsubsection{Photon decoupling}

The partition function of the theory, 
\beqn
Z = \int \cD A  \hskip 2mm \sumint{{}_\mon}  e^{ - S[A,\varrho]}
\label{eq:Z:0}
\eeqn
involves the integration over all photon configurations and over all monopole (and anti-monopole) configurations. These configurations are represented by the monopole density $\varrho = \varrho({\bs x})$. The monopole integration measure 
\beqn
\sumint{{}_\mon} = \sum_{N=0}^\infty \frac{1}{N!} \prod_{a=1}^N \left(\sum_{q_a = \pm 1} {\bar \zeta} \int d^3 x_a \right).
\label{eq:sumint}
\eeqn
is given by the sum over the total number of monopoles $N$, the integration over the positions $\bx_a$ of all these $N$ monopoles, and the sum over all their magnetic charges $q_a = \pm 1$. We work  in the dilute monopole gas regime where overlaps between these magnetic instantons are rare and can therefore be neglected. The factor $1/N!$ takes into account the monopole degeneracy. 

The fugacity parameter $\bar \zeta$ in the measure~\eq{eq:sumint} controls the monopole density. The bar over this quantity implies that this is a bare parameter which is renormalized by interactions.

Due to the decoupling of the photon and monopole parts of the action~\eq{eq:S:A:rho}, the photon and monopole contributions to the partition function~\eq{eq:Z:0} can be factorized:
\beqn
Z & = & Z_\ph \cdot Z_\mon\,, 
\label{eq:Z} \\
Z_\ph & = & \int \cD A \, e^{ - S_\ph[A]}\,, 
\label{eq:Z:ph}\\
Z_\mon & = & \hskip 3mm \sumint{{}_\mon}  e^{ - S_\mon[\varrho]}\,.
\label{eq:Z:mon}
\eeqn
The photon and monopole actions are given in Eqs.~\eq{eq:S:A} and \eq{eq:S:rho}, respectively. In the rest of the paper, we neglect the perturbative physics of photons which does not contribute to the confinement phenomenon.

\subsection{Monopole dynamics}

\subsubsection{Screening and confinement}

Using the explicit expression for the monopole density~\eq{eq:rho}, one can rewrite the monopole action~\eq{eq:S:rho} in terms of the monopole Coulomb gas:
\beqn
S^\mon[\varrho] = \frac{g^2_\mon}{2} \sum_{\stackrel{{a,b=1}}{a \neq b}}^N q_a q_b \, G({\bs x}_a-{\bs x}_b) + S^\mon_0\,.
\label{eq:S:Coulomb:gas}
\eeqn
The action contains the divergent term 
\beqn
S^\mon_0 = \frac{g^2_\mon}{2} \sum_{a=1}^{\infty} G({\bs x}_a,{\bs x}_a)\,,
\label{eq:S0}
\eeqn
which corresponds to the monopole self-energy. It will be renormalized below. In the unbounded $\R^3$ spacetime, the local part of the monopole action~\eq{eq:S0} is a translationally invariant quantity. Below we will consider the theory in a spatially bounded domain, where the monopole self-action~\eq{eq:S0} becomes an explicitly coordinate-dependent quantity.

The monopole partition function~\eq{eq:Z:mon}, can be reformulated in terms of the sine-Gordon model~\cite{Polyakov:1976fu}. Adapting the standard derivation to a finite domain, we get the following equivalence:
\beqn
Z_\mon & = & \sum_{N=0}^\infty \frac{1}{N!} \prod_{a=1}^N \left(\sum_{q_a = \pm 1} \int d^3 x_a \, \zeta({\bs x}_a) \right)  \nonumber \\
& \times & \exp\left[- \frac{g^2_\mon}{2} \int d^3 x \, \int d^3 y \, \varrho(x) D(x-y) \varrho(y)\right] \nonumber \\
& = & \int \cD \chi \exp\left\{ - \int d^3 x \, \cL_{s}(\chi) \right\} \,,
\label{eq:Z:mon:derivation:e}
\eeqn
where
\beqn
\cL_s = \frac{1}{2 g^2_\mon} \bigl[\partial_\mu \chi({\bs x}) \bigr]^2 - 2 \zeta({\bs x}) \cos \chi({\bs x})\,,
\label{eq:cL:sine}
\eeqn
is the Lagrangian of a modified sine-Gordon model. In a contrast to the standard sine-Gordon model, in Eq.~\eq{eq:cL:sine} the fugacity parameter $\zeta$ acquires a dependence on the coordinate $\bs x$ because the divergent monopole self-action~\eq{eq:S0} renormalizes the fugacity parameter
\beqn
\zeta({\bs x}) = {\bar\zeta}\, \exp \left\{ - \frac{2 \pi^2}{g^2} G({\bs x}, {\bs x}) \right\},
\label{eq:zeta:ren}
\eeqn
where we used the identity:
\beqn
{\bar \zeta} e^{-S^\mon_0} = \prod_{a}^N \zeta({\bs x}_a).
\eeqn

In the dilute gas approximation, the vacuum expectation value of the monopole density is proportional to the fugacity parameter:
\beqn
\varrho_\mon({\bs x}) \equiv \biggl\langle \sum_a \delta^{(3)} \left(\bx - \bx_a\right) \biggr\rangle = 2 \zeta({\bs x})\,.
\label{eq:density}
\eeqn
This quantity is the density of monopoles and anti-monopoles, counted as a total number that ignores the magnetic charge. A difference with the definition of the charge-sensitive monopole density~\eq{eq:rho} should be noted.

The presence of the monopole Coulomb gas leads to a screening effect which weakens all interactions at the monopole Debye length $\lambda_D$. At the distance $r$, the interactions are screened exponentially, $e^{- r/\lambda_D}$. The screening length $\lambda_D$ corresponds to the mass $m_\ph$ of the dual field $\chi$ which may be deduced from the dual Lagrangian~\eq{eq:cL:sine}:
\beqn
m_\ph({\bs x}) \equiv \frac{1}{\lambda_D({\bs x})} = \frac{2\pi \sqrt{\varrho_\mon({\bs x})}}{g} = \frac{2\pi \sqrt{2 \zeta({\bs x})}}{g}\,.
\qquad
\label{eq:m:ph}
\eeqn
Here we have used Eq.~\eq{eq:g:m} and expanded the sine-Gordon Lagrangian~\eq{eq:cL:sine} over small fluctuations of the dual field~$\chi$:
\beqn
\cL_s = \frac{1}{2 g^2_\mon} \left[\bigl(\partial_\mu \chi \bigr)^2 + m^2_\ph({\bs x}) \, \chi^2 \right] + O(\chi^4) \,.
\label{eq:cL:sine:exp}
\eeqn
The photon mass~\eq{eq:m:ph} determines the interaction range in the theory. 

Another important property of the compact electrodynamics is the linear confinement of electric charges: a pair of static, electrically charged particle and antiparticle separated by sufficiently large distance $R \gg \lambda_D$ experiences the confining potential $V(r) = \sigma r$ which grows linearly with the distance $R$. This phenomenon exists in the non-Abelian gauge theories, and it is successfully modeled by the compact electrodynamics which is one of a few field-theoretical models where the linear confinement property may be proved analytically. In the latter case, the string tension $\sigma$ is given by the following formula~\cite{Polyakov:1976fu}
\beqn
\sigma({\bs x}) = \frac{8 \sqrt{2 \zeta({\bs x})}}{g_\mon} \equiv \frac{4 g \sqrt{2 \zeta({\bs x})}}{\pi},
\label{eq:string:tension}
\eeqn
which was adapted to the space-dependent fugacity $\zeta$. Due to the non-local character of the string tension, Eq.~\eq{eq:string:tension} is valid for a slowly varying fugacity, where the variation of fugacity at one Debye length is small, $\lambda_D |{\bs\nabla} \zeta({\bs x})| \ll \zeta({\bs x})$.

It is the monopole degree of freedom which leads to the confinement of electric charges: in the absence of monopoles (i.e., if $\varrho_{\mon} = 0$), the effect of confinement disappears. Thus, in our paper, we will study the confining phenomenon in the rotating cU(1) gauge theory concentrating on the properties of the monopoles that cause confinement.

\subsubsection{Deconfinement at finite temperature}

The screening and confinement phenomena, discussed in the previous section, we formulated at zero temperature where all monopoles are not bounded into magnetically neutral pairs. This picture changes at finite temperature, which favors the monopole--anti-monopole binding and, therefore, leads to the deconfinement phenomenon.

At finite temperature $T$, the unbounded Euclidean spacetime $\R^3$ is compactified into $\R^2 \times S^1$, and the finite-temperature Green's function takes the following form:
\beqn
G_T({\bs x}, {\bs x}') = \sum_{l \in \Z} G\biggl({\bs x}, {\bs x}' + \frac{l}{T} \, {\bs{\mathrm{e}}_\tau}\biggr),
\label{eq:G:T}
\eeqn
where the sum goes along the third (imaginary time) axis with the unit vector ${\bs{\mathrm{e}}_\tau}$. The Green's function is, evidently, periodic along this direction:
\beqn
G_T\biggl({\bs x} {+} \frac{m}{T} \, {\bs{\mathrm{e}}_\tau}, {\bs y}\biggr) {=}
G_T\biggl({\bs x}, {\bs y} {+} \frac{m'}{T} \, {\bs{\mathrm{e}}_\tau}\biggr) {=}
G_T({\bs x}, {\bs y}), \quad\quad
\label{eq:G:periodicity}
\eeqn
where $m, m' \in \Z$.

The Green's function~\eq{eq:D} may be written either via the familiar sum over the Matsubara frequencies $\omega_n = 2 \pi n T$ or via the sum in the real spacetime~\eq{eq:G:T}:
\beqn
G_T({\bs x}) & = & T \sum_{n\in \Z} \int \frac{d^2 p}{(2 \pi)^2} \frac{e^{i \omega_n \tau + i {\vec p} {\vec \rho}}}{{\vec p}^{\,2} + \omega^2_n} \nonumber \\
& = & \sum_{l \in \Z} \frac{1}{4 \pi \sqrt{ {\vec \rho}^{\,2} + \left(\tau - \frac{l}{T}\right)^2}}\,.
\label{eq:D:0}
\eeqn

In the high-temperature limit, $T |{\vec \rho}| \gg 1$, the function~\eq{eq:D:0} may be evaluated analytically. In the second line of Eq.~\eq{eq:D:0}, we replace the sum over $l$ via an integral over $l$ and absorb the time variable, $l \to l + T \tau$. Then we notice that while the integral is logarithmically divergent, the divergent part does not depend on the coordinate $\vec x$. After a regularization, we get the finite part:
\beqn
G^{2d}_T({\vec \rho}) = - \frac{T}{2\pi}  \ln T |{\vec \rho}|,
\label{eq:log:0}
\eeqn
where the subleading corrections in the limit $T |{\vec \rho}|  \to 1$ are not shown.

Therefore, at finite temperature the space-dependent part of the monopole action~\eq{eq:S:Coulomb:gas},
\beqn
S^\mon[\varrho] =  \frac{g^2_\mon}{2} \sum_{\stackrel{{a,b=1}}{a \neq b}}^N q_a q_b \,  G^{2d}_T({\vec \rho}_a - {\vec \rho}_b),
\label{eq:S:mon:T}
\eeqn
takes the following form:
\beqn
S^\mon[\varrho] = -  \frac{T g^2_\mon}{4\pi} \sum_{\stackrel{{a,b=1}}{a \neq b}}^N q_a q_b \, \ln T|{\vec \rho}_a- {\vec \rho}_b|.
\label{eq:S:Coulomb:gas:T}
\eeqn
It implies that at large distances $|{\vec \rho}_a - {\vec \rho}_b| \gg 1/T$, a monopole ($q_a = + 1$) and an anti-monopole ($q_b = -1$) experience a logarithmically bounding potential.
Thus, at large temperatures, the logarithmic potential becomes strong and the monopoles and anti-monopoles tend to form bound pairs which cannot support the confinement property. The binding effect results in a change of the thermodynamic phase which is known as the Berezinskii-Kosterlitz-Thouless phase transition~\cite{Berezinskii:1971fst,Berezinskii:1971scn,Kosterlitz:1973xp}.\footnote{The presence of matter fields can change the type of the finite temperature deconfining phase transition~\cite{Agasian:1997wv,Dunne:2000vp}.}

There is a simple way to estimate the transition temperature. Let us consider a single monopole pair (${M\overline{M}}$) at the distance $|{\vec \rho}_a- {\vec \rho}_b|= r$. For a moment, let's consider a discretized space with the lattice spacing $a$ (the smallest distance between the nearest points) and of the same size $L$ in $x$ and $y$ directions such that $a \ll r \ll L $. Therefore, the number of all possible states of a pair ${M\overline{M}}$ pair of the size $r$ is approximately 
\beqn
N(R) = \left(\frac{L}{a}\right)^2 \times \frac{2 \pi r}{a}, 
\label{eq:N:R}
\eeqn
where the first term counts the number of possible positions of the monopole and the second term takes into account the all possible orientations of the monopole with respect to the fixed antimonopole. The boundary terms are proportional to the number of the perimeter states $L/a$, they give a subleading correction to \eq{eq:N:R} linear in $L$, and therefore they are neglected. Thus, the contribution of ${M\overline{M}}$ the pair of radius $r$ to the monopole partition function is therefore
\beqn
\cZ^\mon(R) & = & C_1 N(R) \, e^{-S_{M\overline{M}}}  \nonumber \\
& = & C_2 \frac{R}{a} \exp \left[ - \frac{T g^2_\mon}{4\pi} \ln \frac{r}{a} \right] \nonumber \\
& = & C_2 \exp \left[ \left( 1 - \frac{T g^2_\mon}{4\pi}  \right)\ln \frac{r}{a} \right],
\label{eq:Z:mon:R}
\eeqn
where the action of the single ${M\overline{M}}$ pair can be deduced from the total action \eq{eq:S:Coulomb:gas:T}. The constants $C_1$ and $C_2$ do not play any role in our derivation.

Equation~\eq{eq:Z:mon:R} shows that the contribution of the energy and the entropy has the same functional form. Both contributions cancel each other at 
\beqn
T_c = \frac{4 \pi}{g_\mon^2} \equiv \frac{g^2}{\pi},
\label{eq:Tc:inf}
\eeqn
where we took into account Eq.~\eq{eq:g:m}. At $T > T_c$, the large distances $r$ are unfavorable and therefore the monopoles and anti-monopoles are bound into the ${M\overline{M}}$  pairs. The confinement is lost at high temperatures. At $T<T_c$ the pairs may be of any size~\eq{eq:Z:mon:R} thus signaling the validity of the confinement regime.

Notice that the critical temperature~\eq{eq:Tc:inf} does not depend on the monopole density $\varrho_\mon$ (or, equivalently, on the fugacity $\zeta$). This property is easy to understand on physical grounds. At the critical temperature, all monopoles get bounded into neutral pairs, and this fact does not depend how many monopoles were originally in the confinement phase (here we work in the dilute gas approximation what neglects nonlinear effects). A large number of monopoles in the confinement phase will be reduced to a large number of dipoles, that will not be able to support confinement anyway.

\subsubsection{Applicability}

The phenomena discussed in our work are described in the dilute gas approximation. In this regime, the monopole density is much smaller than the characteristic mass scale determined by the gauge coupling constant:
\beqn
\varrho^{1/2}_\mon \ll \frac{g^3}{(2\pi)^3}, 
\qquad\ \mbox{or} \qquad\ 
\varrho^{1/2}_\mon \; g_\mon^3 \ll 1\,.
\label{eq:applicability}
\eeqn
In the next section, we study the properties of the Green's function $G({\bs x}, {\bs x}')$  in the cylindrical spacetime suitable for the description of the rotating thermal state. This Green's function determines the monopole action~\eq{eq:S:rho}, the screening mass~\eq{eq:m:ph}, and the confining and deconfining properties of the theory.

It is important to stress that across the deconfinement transition, the total density of monopoles and anti-monopoles does not experience any change. The transition is of a geometrical type: these individual objects get combined into pairs while the total number of monopoles and anti-monopoles does not change. In terms of the thermodynamic quantities, the finite-temperature transition in cQED is very smooth and, therefore, it is often said, that the transition is of an infinite order. 

Obviously, the confining features of the system are defined only by the density of the individual monopoles and anti-monopoles that are unbounded into magnetically neutral pairs (we refer to Refs.~\cite{Chernodub:2000mi,Chernodub:2001ws} for a detailed discussion).

\section{Green's functions}
\label{sec:greens}

The important role in determining the dynamics of any quantum system is played by two-point correlations functions which define the propagation of physical degrees of freedom in the system. In our case, the two-point function $G({\bs x}, {\bs x'}) $ enters the monopole action~\eq{eq:S:rho}.  In the dilute gas regime, one may safely neglect quantum loop corrections to the propagator $G$ and treat it as the tree-level Green's function of the differential operator~\eq{eq:Delta:G}. The simplicity of the Laplacian operator is well compensated by the complicated nature of the environment where the Green's function needs to be evaluated: we need to know $G({\bs x}, {\bs x'})$ at finite temperature (with periodic boundary conditions in the imaginary time direction) subjected to a uniform rotation in the spatially cylindrical domain with the spatial Dirichlet boundary conditions.

The finite-temperature Green's functions in rotating systems have been discussed in Ref.~\cite{Vilenkin:1980zv}. Here we choose another method to address this problem: instead of presenting the Green's function as a series over the appropriate eigenfunctions, we use a trick of mirror images to simplify the expressions.  The method of images has been successfully used in solving electrostatic problems of classical electrodynamics~\cite{ref:images} (as brief pedagogical introduction to the subject may be found in Ref.~\cite{ref:quick}). For practical reasons, we consider 2+1 dimensional Green's functions in a disk geometry which are relevant to the problem of confinement in 2+1 dimensional compact electrodynamics. The generalization to higher dimensions is straightforward.

\subsection{Green's function inside a cylinder.} 

First, we consider a cylinder of the radius $R$ in the $(x_1,x_2)$ plane with the center at the origin $x_1 = x_2 = 0$, and with the axis along the $\tau \equiv \tau$ direction of an infinite height. We denote 
\beqn
{\bs x} = ({\vec \rho},\tau) \quad \mbox{with} \quad {\vec \rho} = (x_1,x_2).
\eeqn

We consider the symmetric Green's function, 
\beqn
G({\bs x}, {\bs x'}) = G({\bs x'}, {\bs x}), 
\label{eq:G:symmetry}
\eeqn
of the Laplacian:
\beqn
\Delta^{(3)}_x = \frac{\partial^2}{\partial x_1^2} + \frac{\partial^2}{\partial x_2^2} + \frac{\partial^2}{\partial \tau^2},
\eeqn
satisfying the equation
\beqn
\Delta_x G({\bs x}, {\bs x'}) {\biggl|}_{|{\vec \rho}|,|{\vec \rho}{\,}'| < R}= - \delta({\bs x} - {\bs x'}),
\label{eq:Greens}
\eeqn
and obeying the Dirichlet boundary condition:
\beqn
G({\bs x}, {\bs x'}) {\biggl |}_{|{\vec \rho}| = R} = G({\bs x}, {\bs x'}) {\biggl |}_{|{\vec \rho}{\,}'| = R} = 0.
\label{eq:G:Dirichlet}
\eeqn
The Green's function is:
\beqn
G({\bs x}, {\bs x'}) = \frac{1}{4 \pi} \left( \frac{1}{|{\bs x} - {\bs x'}|} - \frac{1}{|{\bs x} - {\bs x'}|_*} \right),
\label{eq:G}
\eeqn
where
\beqn
|{\bs x} - {\bs x'}| & = & \sqrt{ | {\vec \rho} -  {\vec \rho}{\,}'|^2 + (\tau - \tau')^2}, \label{eq:xy:1} \\
|{\bs x} - {\bs x'}|_* & = & \left[ | {\vec \rho} -  {\vec \rho}{\,}'_*|^2 \, |{\vec \rho}{\,}'|^2 \frac{1}{R^2} + (\tau - \tau')^2 \right]^{\frac{1}{2}}.
\label{eq:xy:2} 
\eeqn
The second term in Eq.~\eq{eq:G} involves the exterior image
\beqn
{\bs x}_* = ({\vec \rho}_*, \tau) \quad \mbox{with} \quad {\vec \rho}_* \equiv {\vec \rho}_*\left( {\vec \rho} \right)= \frac{R^2}{|{\vec \rho}|^2} \, {\vec \rho},
\label{eq:mirror:point}
\eeqn
of the interior point ${\bs x}$ mirrored in the reflective surface of the cylinder. 

The Green's function~\eq{eq:G} satisfies the boundary condition~\eq{eq:G:Dirichlet} because of the following relation for the image point~\eq{eq:mirror:point}: 
\beqn
{\vec \rho}_* (|{\vec \rho}| = R) = {\vec \rho}.
\label{eq:x:star:R}
\eeqn 
Using the identity 
\beqn
| {\vec \rho} -  {\vec \rho}{\,}'_*|^2 \, |{\vec \rho}{\,}'|^2 = | {\vec \rho}_* -  {\vec \rho}{\,}'|^2 \, |{\vec \rho}|^2 \equiv | {\vec \rho}{\,}' -  {\vec \rho}_*|^2 \, |{\vec \rho}|^2,
\qquad
\eeqn
one can also show that the Green's function~\eq{eq:G} is a symmetric function of its variables~\eq{eq:G:symmetry} despite a seemingly asymmetric representation of the distances~\eq{eq:xy:1} and \eq{eq:xy:2}. In the large-radius limit $R \to \infty$, the Green's function reduces to the standard expression with the second term in Eq.~\eq{eq:G} vanishing.

Similarly to the Green's function on $\R^3$, Eq.~\eq{eq:D}, the Green's function inside the cylinder~\eq{eq:G} may also be rewritten as the integral over the momentum ${\bs p} = ({\vec p}, p_\tau)$:
\beqn
G({\bs x}, {\bs x}') & = & \int \frac{d^3 p}{(2\pi)^3} \frac{e^{i p_\tau (\tau - \tau')}}{{\vec p}^{\,2} + p_\tau^2} \nonumber \\
& & \times \left( e^{i {\vec p} \cdot ({\vec \rho}- {\vec \rho})} - e^{i {\vec p} \cdot  ({\vec \rho}- {\vec \rho}{\,}'_*) |{\vec \rho}{\,}'|/R} \right),
\label{eq:G:T:0}
\eeqn
In is important to notice that the two-dimensional transverse momentum ${\vec p}$ is not quantized and the integration goes over the unbounded region in the momentum space. This is not an error since the Dirichlet boundary condition is enforced by the unconventional exponential factors in Eq.~\eq{eq:G:T:0}.

\subsection{Finite-temperature Green's function in a disk} 

As we mentioned above, at a finite temperature $T>0$, the symmetry axis $\tau$ of the cylinder is compactified to a circle of the length $1/T$ closed via the periodic boundary condition. The finite-temperature Green's function takes the generic form of a sum along the height of the compactified cylinder~\eq{eq:G:T}. 
The finite-temperature function may also be rewritten, similarly to the $T=0$ representation~\eq{eq:G:T:0}, in terms of a sum in the momentum space over the bosonic Matsubara frequencies $\omega_n = 2 \pi T n$:
\beqn
G_T({\bs x}, {\bs x}') & = & T \sum_{n \in \Z} \int \frac{d^2 p}{(2\pi)^2} \frac{e^{i \omega_n (\tau - \tau')}}{{\vec p}^{\,2} + \omega_n^2} \nonumber \\
& & \times \left( e^{i {\vec p} \cdot ({\vec \rho}- {\vec \rho})} - e^{i {\vec p} \cdot  ({\vec \rho}- {\vec \rho}{\,}'_*) |{\vec \rho}{\,}'|/R} \right),
\label{eq:G:T:Matsubara}
\eeqn
where the dot denotes the scalar product in two-dimensions, ${\vec p} \cdot {\vec \rho} = p_1 x_1 + p_2 x_2$. Notice that rotation has not been implemented yet.

In a case when the third coordinates of the points ${\bs x}$ and ${\bs y}$ coincide, $\tau = y_3$ [modulo the periodic shits~\eq{eq:G:periodicity}], the propagator~\eq{eq:G:T} takes a bit more familiar form:
\beqn
G_T({\vec \rho}, {\vec y}) & \equiv & G_T({\bs x}, {\bs y}){\biggl|}_{\tau = y_3} = \int \frac{d^2 p}{(2 \pi)^2}  \frac{1}{p}
\label{eq:G:T:neq0}\\
& & \hskip -7mm \times \left( e^{i {\vec p} \cdot ({\vec \rho}- {\vec y})} - e^{i {\vec p} \cdot ({\vec \rho}- {\vec y}_*) |\vec y|/R} \right) \left( \frac{1}{2} + f_T(p) \right),
\nonumber 
\eeqn
with the Bose-Einstein distribution function
\beqn
f_T(p) = \frac{1}{e^{p/T} - 1}.
\label{eq:f:T}
\eeqn

\subsection{Green's function in a rotating disk at finite temperature} 

\subsubsection{Derivation}

Let us come back to the momentum representation of the Green's function~\eq{eq:G:T:Matsubara}. The integral over the two-dimensional momentum ${\vec p}$ may be represented in two equivalent ways which include either the integral over the angular variable of the momentum $\theta_p$ or the sum over the associated quantized angular momentum $m$. Let us denote  ${\vec p} = (p \cos \theta_p, p \cos \theta_p) $ and  ${\vec \xi} = (\xi \cos \theta_\xi, \xi \cos \theta_\xi) $ with 
$p = |{\vec p}|$ and $\xi = |{\vec \xi}|$. Here ${\vec \xi}$ is a vector in the two-dimensional plane (either ${\vec \xi} = {\vec \rho}$ or ${\vec \xi} = {\vec y}$). Then the integral over the momentum may be written as follows:
\beqn
\int \frac{d^2 p}{(2\pi)^2} e^{i {\vec p} \cdot {\vec \xi}} Q(p) & = & \int\limits_0^\infty p d p \int\limits_0^{2\pi} \frac{d \theta_p}{(2 \pi)^2} e^{i p \xi \cos(\theta_p - \theta_\xi)} Q(p)\quad\nonumber \\
& = & \int\limits_0^\infty \frac{p d p}{2 \pi} J_{0} (p \xi) Q(p),
\eeqn
where $J_l(x)$ is the Bessel function of zero order, $l = 0$, and $Q(p)$ is an artbitrary function of the absolute value of the momentum~$p$.

The integral over the angle $\theta_p$ in the momentum space may also alternatively be rewritten as a sum over the quantized angular momentum $m$:
\beqn
\int\limits_0^{2\pi} \frac{d \theta_p}{(2 \pi)^2}  e^{i {\vec p} \cdot ({\vec \rho}- {\vec y})} = \sum_{m \in \Z} \phi_{p,m}({\vec \rho}) \phi^*_{p,m}({\vec y}),
\label{eq:identity:1}
\eeqn
in terms of the eigenfunctions 
\beqn
\phi_{p, m}({\vec \xi}\,) = \frac{1}{\sqrt{2\pi}} e^{i m \theta_\xi } J_m(p \xi),
\label{eq:phi:pm}
\eeqn
of the two-dimensional Laplacian operator 
\beqn
\Delta_{\xi}^{(2d)} = \frac{\partial^2 }{\partial \xi_1^2} + \frac{\partial^2 }{\partial \xi_2^2}.
\eeqn
In the polar coordinates $\xi_1 + i \xi_2 = \xi e^{i \theta_\xi}$, the eigenvalue equation reads as follows:
\beqn
\left( \frac{1}{\xi} \frac{\partial }{\partial \xi} \xi  \frac{\partial }{\partial \xi} + \frac{1}{\xi^2}  \frac{\partial^2 }{\partial \theta_\xi^2}\right) \phi_{p, m}({\vec \xi}\,) = 
- p^2 \phi_{p, m}({\vec \xi}\,).
\eeqn

To derive Eq.~\eq{eq:identity:1}, we implemented the plane wave expansion in two dimensions:
\beqn
e^{i z \cos \theta} = \sum_{m \in \Z} i^m e^{i m \theta} J_m(z).
\label{eq:plane:wave}
\eeqn

Using the properties of the Bessel functions, it is not difficult to show that the wavefunctions~\eq{eq:phi:pm} form a complete set of functions on the disk:
\beqn
& & \sum_{m \in \Z} \int\limits_0^\infty p d p\,  \phi^*_{p, m}({\vec \xi}\,) \phi_{p, m}({\vec \xi}\,') = \frac{\delta(\xi - \xi')}{\xi} \nonumber \\
& & \qquad \times \sum_{l} \delta(\theta_\xi - \theta_{\xi'} - 2 \pi l)  \equiv \delta({\vec \xi} - {\vec \xi}\,'). 
\label{eq:completeness}
\eeqn
In order to find the appropriate Green's function, we do need to ensure the completeness~\eq{eq:completeness} of these auxiliary functions on a disk. As will see below, the orthonormalization properties of the functions~\eq{eq:phi:pm} do not play any role in our derivation.

The choice of the representation is a matter of convenience. Below we pass to the angular momentum representation of the integration measure:
\beqn
\int \frac{d^2 p}{(2\pi)^2} \to \sum_{m \in \Z} \int\limits_0^\infty \frac{p d p}{2\pi},
\eeqn
Specifically, the Green's function~\eq{eq:G:T:Matsubara} can now be identically rewritten as follows:
\beqn
& & G_T({\bs x}, {\bs x}') = T \sum_{n \in \Z} \sum_{m \in \Z} \int\limits_0^{\infty} p d p \frac{e^{i \omega_n (\tau - \tau')}}{p^2 + \omega_n^2} 
\label{eq:G:T:Matsubara:m} \\
& & \times  \left[ \phi_{p,m}({\vec \rho}) \phi^*_{p,m}({\vec \rho}{\,}') - \phi_{p,m}\left(\frac{{\vec \rho} |{\vec \rho}{\,}'|}{R}\right) \phi^*_{p,m}\left(\frac{{\vec \rho}{\,}'_* |{\vec \rho}{\,}'|}{R}\right) \right]
\nonumber
\eeqn
As we discussed earlier, the inclusion of the rotation can be taken into account via the shift of the Matsubara frequencies: $\omega_n \to \omega_n - i \Omega m$.
Performing this operation in Eq.~\eq{eq:G:T:Matsubara:m}, we arrive to the following representation of the Green's function of the Laplacian on a spatial disk of the radius $R$ in the reference frame rotating with the angular velocity $\Omega$ at the finite temperature $T$:
\beqn
& & G_{T,\Omega}({\bs x}, {\bs x}') = T \sum_{n \in \Z} \sum_{m \in \Z} \int\limits_0^{\infty} p d p \frac{e^{i \omega_n (\tau - \tau')}}{p^2 + (\omega_n - i \Omega m)^2} 
\label{eq:G:T:Omega} \\
& & \times  \left[ \phi_{p,m}({\vec \rho}) \phi^*_{p,m}({\vec \rho}{\,}') - \phi_{p,m}\left(\frac{{\vec \rho} |{\vec \rho}{\,}'|}{R}\right) \phi^*_{p,m}\left(\frac{{\vec \rho}{\,}'_* |{\vec \rho}{\,}'|}{R}\right) \right],
\nonumber
\eeqn
where $\phi_{p,m}({\vec \rho})$ are the eigenfunctions~\eq{eq:phi:pm} of the two-dimensional Laplacian.

\subsubsection{Properties}

After performing all these transformations, it is less than obvious that the expression~\eq{eq:G:T:Omega} is indeed the Green's function of the Laplacian in rotating frame which satisfy all necessary requirements that are fulfilled by its counterpart written in the static frame: It should satisfy the appropriate Green's function equation similar to Eq.~\eq{eq:Greens}, be symmetric with respect to the permutation of arguments~\eq{eq:G:symmetry}, and obey the Dirichlet boundary condition at the boundary~\eq{eq:G:Dirichlet}. Below, we will perform a step-by-step check of all these requirements. 

{\bf The main equation}. The Laplacian in the rotating frame in the imaginary time (after performing the Wick rotation) has the following form:
\beqn
\Delta^{(3)}_{\Omega, x} = \frac{\partial^2}{\partial x_1^2} + \frac{\partial^2}{\partial x_2^2} + \left( \frac{\partial}{\partial \tau} - i \Omega \frac{\partial}{\partial \theta } \right)^2.
\label{eq:Delta:rotating}
\eeqn
Taking into account that the wave function
\beqn
\phi_{p,m,n}({\bs x})  = e^{i \omega_n \tau} \phi_{p,m}({\vec \rho})
\eeqn
is the eigenfunction of the Laplacian~\eq{eq:Delta:rotating},
\beqn
\left[ \Delta^{(3)}_{\Omega, x} + p^2 + (\omega_n - i \Omega m)^2 \right] \phi_{p,m,n}({\bs x}) = 0,
\eeqn
we determine that the first term in the square brackets of Eq.~\eq{eq:G:T:Omega} gives us:
\beqn
\Delta^{(3d)}_{\Omega, x}  G_{T,\Omega}({\bs x}, {\bs x}')  = - \delta({\bs x} - {\bs x}'),
\label{eq:delta:G:3}
\eeqn
where the delta function takes the periodicity with respect to the compactified time direction:
\beqn
\delta({\bs x} - {\bs x}') = \delta({\vec \rho} - {\vec \rho}{\,}') \sum_{l} \delta(\tau - \tau' - l/T).
\eeqn
The second term in the square brackets of Eq.~\eq{eq:G:T:Omega} does not contribute to Eq.~\eq{eq:delta:G:3} 
because the points in this term are located at different sides with respect to the disk's boundary.

{\bf The symmetry under the flip of the arguments.} In the first term in the square brackets of Eq.~\eq{eq:G:T:Omega}, the permutation of the arguments is 
equivalent to the flipping of the sign in front of the angular momentum, $m \to - m$:
\beqn
\phi_{p,m}({\vec \rho}{\,}') \phi^*_{p,m}({\vec \rho})  = \phi_{p,-m}({\vec \rho}) \phi^*_{p,-m}({\vec \rho}{\,}').
\eeqn
This fact follows from the definition of the wave function~\eq{eq:phi:pm}.

It is less obvious but easy to show that the same is also true for the second term in the square brackets of Eq.~\eq{eq:G:T:Omega}:
\beqn
& & \phi_{p,m}\left(\frac{{\vec \rho}{\,}' |{\vec \rho}|}{R}\right) \phi^*_{p,m}\left(\frac{{\vec \rho}_* |{\vec \rho}|}{R} \right) \nonumber \\
& & \qquad = \phi_{p,-m}\left(\frac{{\vec \rho} |{\vec \rho}{\,}'|}{R}\right) \phi^*_{p,-m}\left(\frac{{\vec \rho}{\,}'_* |{\vec \rho}{\,}'|}{R} \right),
\eeqn
since the mirror point ${\vec \rho}_*$ defined in Eq.~\eq{eq:mirror:point} has the same angular coordinate, $\theta_{x_*} \equiv \theta $ and satisfies the identity
$|{\vec \rho}_*| |{\vec \rho}| = R^2$. The flip in $m$ is compensated by the change of the angular summation variable $m \to - m$ and a simultaneous flip in the sum $n \to - n$ over the Matsubara frequencies in Eq.~\eq{eq:G:T:Omega}. We thus prove the anticipated symmetry:
\beqn 
G_{T,\Omega}({\bs x}, {\bs x}') = G_{T,\Omega}({\bs x}', {\bs x}).
\label{eq:symm:T:Omega}
\eeqn

{\bf Boundary condition.} The property~\eq{eq:x:star:R} of the mirror point~\eq{eq:mirror:point} implies that if the point ${\vec \rho}{\,}'$ touches the boundary of the disk, $|{\vec \rho}{\,}'| = R$, then the second term in the square brackets of Eq.~\eq{eq:G:T:Omega} becomes equal to the first term and the Green's function~\eq{eq:G:T:Omega} vanishes identically. Due to the symmetry~\eq{eq:symm:T:Omega}, the same statement is valid for the ${\vec \rho}$ point. We arrive to the conclusion that the Green's function~\eq{eq:G:T:Omega} satisfies the required Dirichlet boundary condition:
\beqn
G_{T,\Omega}({\bs x}, {\bs x'}) {\biggl |}_{|{\vec \rho}| = R} = G_{T,\Omega}({\bs x}, {\bs x'}) {\biggl |}_{|{\vec \rho}{\,}'| = R} = 0.
\label{eq:G:Omega:Dirichlet}
\eeqn
Thus, the boundary is a reflecting surface and the particles cannot travel outside the rotating disk.

\subsubsection{Coinciding points}

It is easier to analyze the propagator at coinciding imaginary times, $\tau = \tau'$. In this case, the Green's function~\eq{eq:G:T:Omega} may be rewritten in the following form:
\beqn
& & G_{T,\Omega}({\vec \rho}, {\vec \rho}{\,}') \equiv G_{T,\Omega}({\bs x}, {\bs x}') {\biggl|}_{\tau = \tau'} = 
\frac{1}{2} \sum_{m \in \Z} \int\limits_0^{\infty} d p
\nonumber \\
& & \times \left[ 1 + \frac{1}{e^{(p + \Omega m)/T}-1}  + \frac{1}{e^{(p - \Omega m)/T}-1}  \right]
\label{eq:G:T:Omega:t0} \\
& & \times  \left[ \phi_{p,m}({\vec \rho}) \phi^*_{p,m}({\vec \rho}{\,}') - \phi_{p,m}\left(\frac{{\vec \rho} |{\vec \rho}{\,}'|}{R}\right) \phi^*_{p,m}\left(\frac{{\vec \rho}{\,}'_* |{\vec \rho}{\,}'|}{R}\right) \right],
\nonumber
\eeqn
where we used the summation identity~\eq{eq:S:b}:
\beqn
& & T \sum_{n\in\Z} \frac{1}{\left( \omega_n - i \Omega m \right)^2 + p^2} \\
& & \hskip 5mm = \frac{1}{2 p}  \left[ 1 + \frac{1}{e^{(p + \Omega m)/T}-1}  + \frac{1}{e^{(p - \Omega m)/T}-1}  \right], \nonumber
\eeqn
with $p \equiv |{\vec p}|$. Due to the translational invariance, the Green's function~\eq{eq:G:T:Omega:t0} does not depend on imaginary time. We also notice in Eq.~\eq{eq:G:T:Omega:t0} two Bose-Einstein distribution functions~\eq{eq:f:T} with the energies modified by rotation.

At the coinciding spatial points, understood as a limit ${\vec \rho}{\,}' \to {\vec \rho}$ (or, equivalently, as ${\bs x}' \to {\bs x}$), we get:
\beqn
& & G_{T,\Omega}(\rho) \equiv G_{T,\Omega}({\bs x}, {\bs x}') {\biggl|}_{\stackrel{{\bs x}' \to {\bs x}}{\tiny{|{\bs x}| = \rho}}} = 
\sum_{m \in \Z} \int\limits_0^{\infty} \frac{d p}{4 \pi} 
\nonumber \\
& & \times \left( 1 + \frac{1}{e^{(p + \Omega m)/T} -1}  + \frac{1}{e^{(p - \Omega m)/T}-1}  \right)
\label{eq:G:T:Omega:x0} \\
& & \times  \left[ J^2_m(p \rho) - J_m(p R) J_m\left( \frac{p \rho^2}{R} \right) \right],
\nonumber
\eeqn
where we used Eq.~\eq{eq:phi:pm} and denoted the radial variable $\rho = |{\vec \rho}|$. The expression~\eq{eq:G:T:Omega:x0} is formally divergent and therefore an ultraviolet regularization of the integral and the sum are implicitly assumed (we will consider the regularization shortly below). Finally, it is worth reminding that Eq.~\eq{eq:G:T:Omega:x0} is valid only inside the disk, at $\rho \leqslant R$.

From the form of Eq.~\eq{eq:G:T:Omega:x0}, we can immediately make two conclusions. First, the particle in the very center is not affected by the rotation since 
$G_{T,\Omega}(0)  = G_{T}(0)$. Indeed, the Bessel function vanish at vanishing arguments unless $m=0$. At $m=0$, however, the Green's function loses the dependence on the rotation frequency $\Omega$. Second, the Green's function vanishes at the boundary, $G_{T,\Omega}(R) = 0$, as expected.

\subsubsection{Causality}

The causality of a rigidly rotating disk is maintained provided the angular frequency is bounded, $|\Omega| R < 1$, so that the boundary of the disk rotates with velocity smaller than the speed of light. Since the particle cannot escape the disk, the causality property is maintained. 

However, one may still question the validity of the causality for the proposed representation of the propagator because we have chosen the representation~\eq{eq:G:T:Omega} in which the radial momentum is not quantized. While this non-quantization property is a technically convenient feature of our representation, it puts a shadow on the validity of the causality principle in our solution as the non-quantization of the transverse momentum is usually associated with unbounded, and thus causality-violating, when rigidly rotating, domains. 

The potential violation of the causality could be well guessed in Eq.~\eq{eq:G:T:Omega:x0}. To this end, let us assume that the factors in the round brackets are associated with the Bose-Einstein occupation numbers which may become negative provided $p \pm \Omega m < 0$. This would-be unphysical behavior does not appear in a traditional approach where the momenta $p = p_{ml}$ are quantized and $p_{ml} \pm \Omega m > 0$ for any angular $m$ and radial $l$ quantum numbers.

The causality violation leads to an accumulation of an infinite tower of the states with a negative occupation numbers. For example, for a clockwise rotation ($\Omega > 0$), this dangerous tower builds up at a fixed $p$ and sufficiently large positive $m$:
\beqn
f_T(p,) = \frac{1}{e^{(p \pm \Omega m)/T} - 1} < 0 \quad \mbox{for} \quad m > p/\Omega. \qquad
\label{eq:f:T:negative}
\eeqn

In our representation, the unphysical, negative-occupation tower does not exist due to the presence of the second, image-mirrored term in propagator~\eq{eq:G:T:Omega}. It's role is well seen in Eq.~\eq{eq:G:T:Omega:x0}. To this end we use the large-order representation of the Bessel functions,
\beqn
J_m(z) = \frac{1}{\sqrt{2 \pi m}} \left( \frac{e z}{2 m} \right)^m + \dots,
\eeqn
where we consider a finite argument $z$ and take $m$ to be positive real values (similar arguments hold for negative values as well). Substituting this expansion in Eq.~\eq{eq:G:T:Omega:x0}, we see that the leading terms in the square brackets exactly cancel each other and the negative tower never builds up. Moreover, the problematic negative last term in the round brackets cancels with the first term, thus leaving only the first Bose-Einstein term with the correct, positive occupation number. The same is obviously true for the counterclockwise rotation $\Omega < 0$, where the Bose-Einstein terms exchange their roles.

This result can be understood on physical grounds: the causality cannot be broken since the interior and exterior of the disk do not communicate with each other while every point of the disk rotates with the velocity lower than the speed of light ($|\Omega| R < 1$). Thus, we refute the suspected causality violation in our representation~\eq{eq:G:T:Omega} of the finite-temperature Green's function in the rotating frame by highlighting the important role that is played by the second (image mirror) term in the square brackets of Eq.~\eq{eq:G:T:Omega}. The presence of this term restores the causality of the Green's function. Our result is in line with the general proof that the Dirichlet boundary condition for a scalar field on a surface with $|\Omega| R < 0$ always gives $p \pm \Omega m > 0$~\cite{Nicolaevici:2001yy}.

\subsubsection{Physical Green's function: Can cold vacuum rotate?}

Based on physical grounds, we expect that the answer to this question is ``no'': the vacuum is a Lorentz-invariant environment which cannot rotate because there is no substance that can be rotated. As a result, we expect that the propagator of physical particle should not depend on the angular frequency at zero temperature and in the absence of matter.~\cite{Chen:2015hfc,Chernodub:2016kxh}. However, one may easily check that our propagator~\eq{eq:G:T:Omega} does depend on $\Omega$ at $T=0$. This property does not mean that the expression is subjected of a mathematical inaccuracy. On the contrary, we have checked that our formula for the Green's function fulfills all the requirements imposed on the latter. A resolution of this issue has a physical rather than mathematical nature. 

A Green's function of any operator is identified up to bilinear combinations of zero modes which are vanishing under the action of this operator. Since our derivation differs from the standard approach which employs the sum over all the modes, the difference between these two approaches appears due to omission of these zero modes. 

Usually, the correct Green's function is selected out of many others on the physical grounds (for example, based on an appropriate behaviour of the Green's function at asymptotically large values of it arguments). In our bounded space, this requirement is no more applicable, and therefore we use the other argument: we require that the physical Green's function is independent of the angular frequency $\Omega$ at zero temperature $T$. 

The physical Green's function, constrained by these conditions, is given by the following formula: 
\beqn
G^{\mathrm{phys}}_{T,\Omega}({\bs x}, {\bs x}') & = & G_{T,\Omega}({\bs x}, {\bs x}') - G_{T=0,\Omega}({\bs x}, {\bs x}') \nonumber \\
& & + G_{T=0,\Omega = 0}({\bs x}, {\bs x}').
\label{eq:G:phys}
\eeqn
with $ G_{T,\Omega}({\bs x}, {\bs x}')$ given in Eq.~\eq{eq:G:T:Omega}.

For the physical definition~\eq{eq:G:phys} of the Green's function, we get, obviously, 
\beqn
\Delta^{(3)}_{\Omega, x}  G^{\mathrm{phys}}_{T,\Omega}({\bs x}, {\bs x}')  = - \delta({\bs x} - {\bs x}'),
\eeqn
in addition to the correct boundary conditions,
\beqn
G^{\mathrm{phys}}_{T,\Omega}({\bs x}, {\bs x}') {\biggl|}_{|{\vec \rho}| = 0} =
G^{\mathrm{phys}}_{T,\Omega}({\bs x}, {\bs x}') {\biggl|}_{|{\vec \rho}{\,}'| = 0} = 0, \quad
\eeqn
the permutational symmetry,
\beqn
G^{\mathrm{phys}}_{T,\Omega}({\bs x}, {\bs x}')  = G^{\mathrm{phys}}_{T,\Omega}({\bs x}', {\bs x}), 
\eeqn
and the correct physical limits:
\beqn
G^{\mathrm{phys}}_{T=0,\Omega}({\bs x}, {\bs x}') & = & G_{T=0,\Omega = 0}({\bs x}, {\bs x}'),\\
G^{\mathrm{phys}}_{T,\Omega=0}({\bs x}, {\bs x}') & = & G_{T,\Omega = 0}({\bs x}, {\bs x}').
\eeqn

For coinciding points, one immediately gets from Eqs.~\eq{eq:G:phys} and \eq{eq:G:T:Omega:x0}:
\beqn
& & G^{\mathrm{phys}}_{T,\Omega}(\rho) = 
\sum_{m \in \Z} \int\limits_0^{\infty} \frac{d p}{4 \pi}  \left[ J^2_m(p \rho) - J_m(p R) J_m\left( \frac{p \rho^2}{R} \right) \right]
\nonumber \\
& & \times \left( 1 + \frac{\Theta(p + \Omega m)}{e^{(p + \Omega m)/T} -1}  + \frac{\Theta(p - \Omega m)}{e^{(p - \Omega m)/T}-1}  \right).
\label{eq:G:phys:x0}
\eeqn

\subsubsection{Faster-than-light rotation}

In order to further cross-check our results, we would like to make sure that the physical Green's function~\eq{eq:G:phys} gives us an unphysical result if the causality condition is violated. To this end, it is sufficient to notice from Eq.~\eq{eq:G:phys:x0} that the integration over the transverse momentum proceeds up to the point $p = |\Omega m|$. Taking, for simplicity, $\Omega > 0$ and $m >0$, we obtain that close to this point in the momentum and at high angular velocities, the Bessel functions contain a simple pole
\beqn
J_{m}(m z) \sim \frac{1}{\sqrt{1 - z^2}}, 
\eeqn
which translates to three singularities in Eq.~\eq{eq:G:phys:x0}:
\beqn
\frac{1}{1 - \Omega^2 \rho^2}, \quad  \frac{1}{\sqrt{1 - \Omega^2 R^2}}, \quad  \frac{1}{\sqrt{1 - \Omega^2 \rho^4/R^2}}.
\eeqn
As expected, these singularities will make the expression~\eq{eq:G:phys:x0} divergent provided the causality condition is broken, $\Omega^2 R^2 > 1$.

\subsubsection{Angular frequency dependence}

The monopole density~\eq{eq:density} depends on the propagator evaluated at coinciding points~\eq{eq:zeta:ren}. In order to figure out the effect of the rotation on the monopole density, we should estimate the dependence of the propagator~\eq{eq:G:phys:x0}  on the angular frequency $\Omega$.  This propagator contains an uninteresting zero-temperature part which does not depend on $\Omega$. The subtraction of this contribution from the full Green's function,
\beqn
{\mathcal G}_{\Omega,T}(\rho) =  G^{\mathrm{phys}}_{T,\Omega}(\rho) - G^{\mathrm{phys}}_{T=0,\Omega = 0}(\rho),
\label{eq:cal:G}
\eeqn
amounts to removing the unity from the round brackets in the propagator~\eq{eq:G:phys:x0} at coinciding points. Notice that according to Eq.~\eq{eq:G:phys:x0}, the propagator~\eq{eq:cal:G} does not depend on the angular frequency $\Omega$ at the axis of rotation, $\rho = 0$.

The quantity~\eq{eq:cal:G} is a sum of a finite and divergent parts, respectively:
\beqn
{\mathcal G}_{\Omega,T}(\rho) = {\mathcal G}^{\mathrm{fin}}_{\Omega,T}(\rho) +  {\mathcal G}^{\mathrm{div}}_{\Omega,T}(\rho).
\label{eq:G:cal:sum}
\eeqn
The finite part has the following form:
\beqn
& &  {\mathcal G}^{\mathrm{fin}}_{\Omega,T}(\rho) = 
\sum_{m = 1}^\infty \int\limits_0^{\infty} \frac{d p}{2\pi} \frac{J^2_m(p \rho) - J_m(p R) J_m\left( p \rho^2/R \right)}{e^{(p + \Omega m)/T} -1} 
\nonumber \\
& & 
+\sum_{m = 1}^\infty \ \int\limits_{\Omega m}^{\infty} \frac{d p}{2\pi} \frac{1}{e^{(p - \Omega m)/T} -1} \biggl[J^2_m(p \rho) - J^2_m(m \Omega \rho)
\nonumber  \\
& &  + J_m(m \Omega R) J_m\left( \frac{m \Omega \rho^2}{R} \right) - J_m(p R) J_m\left( \frac{p \rho^2}{R} \right) \biggr].
\label{eq:G:phys:x1}
\eeqn
where we implied, without loss of generality, $\Omega > 0$.

The second term in Eq.~\eq{eq:G:cal:sum} appears because the massless field theories in 2+1 dimensions contain soft logarithmic divergences, regularized by an infrared cutoff $\Lambda_{\mathrm{IR}}$. It is well-known that in cQED this divergence cannot be removed via, for example, a mass-gap generation typical for an interacting theory. On the contrary, this divergence represents a long-range interaction between the monopoles and its absence signals the loss of the mass-gap generation and the confinement phenomena. 

\begin{figure}[!thb]
\centerline{\includegraphics[scale=0.45, clip=true]{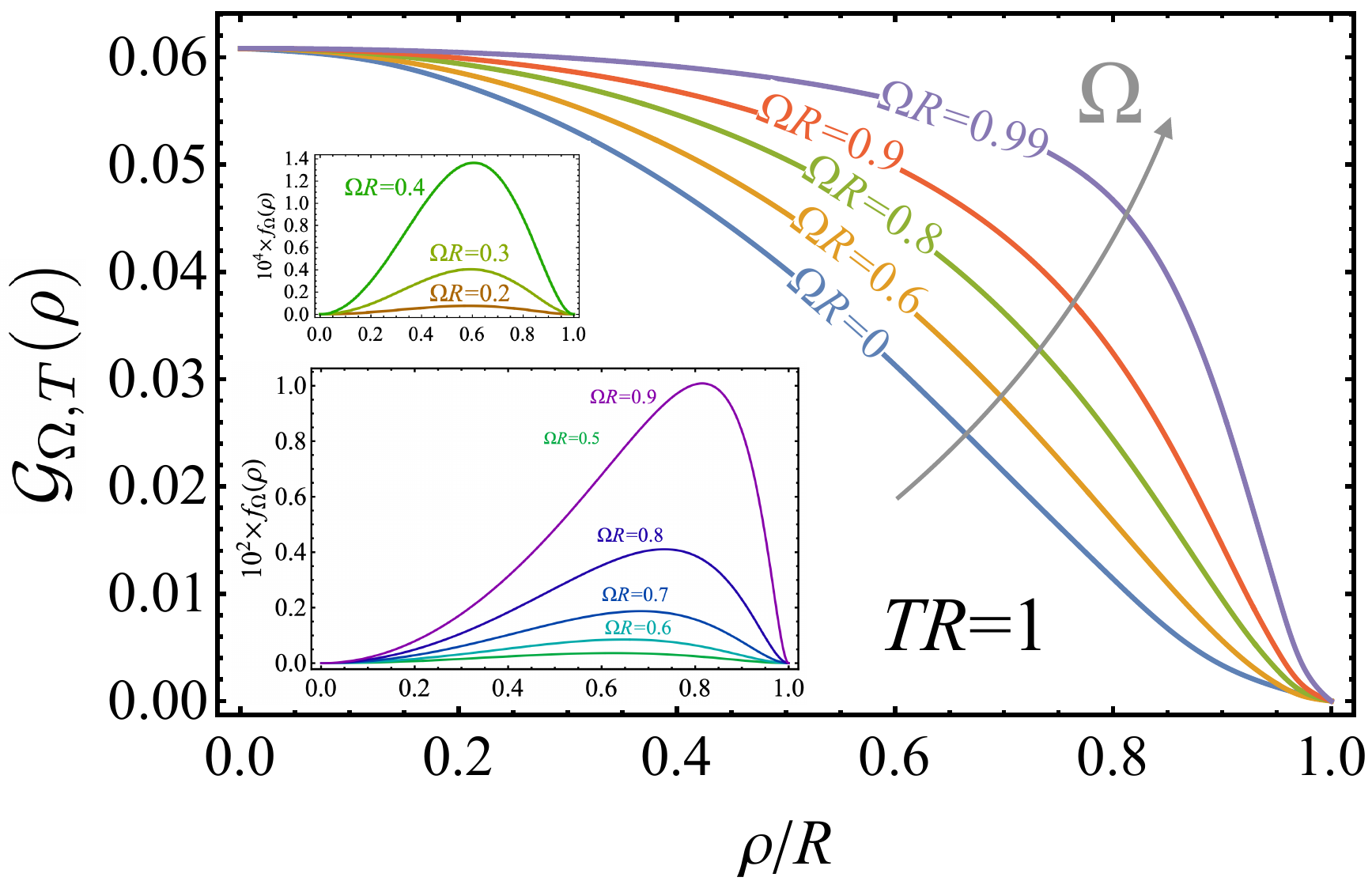}}
\caption{The correction~\eq{eq:cal:G} to the same-point Green's function induced by rotation with the angular frequency $\Omega$ at temperature $T = 1/R$. The insets show the behaviour of the factor $f_\Omega$ given in Eq.~\eq{eq:f:Omega}.}
\label{fig:propagator}
\end{figure}

On physical grounds, we fix the cutoff to the radius of the disk, $\Lambda_{\mathrm{IR}} = 1/R$, and get:
\beqn
{\mathcal G}^{\mathrm{div}}_{\Omega,T}(\rho) = h(T) f_\Omega(\rho),
\eeqn
where
\beqn
h(T) = \frac{T}{2\pi} \log(R T) 
\eeqn
is a temperature prefactor and 
\beqn
f_\Omega(\rho) = \sum_{m = 1}^\infty \biggl[J^2_m(m \Omega \rho) {-} J_m(m \Omega R) J_m\left( \frac{m \Omega \rho^2}{R} \right) \biggr]\!, \qquad
\label{eq:f:Omega}
\eeqn
is a radial factor expressed as the quickly converging sum over the angular momentum~$m$. The dependence of the finite-$\Omega$ correction~\eq{eq:cal:G} to same-point Green's function as well as the behaviour of the factor $f_\Omega$ are shown in Fig.~\ref{fig:propagator}. The response of the Green's function to the rotation is important for the confining properties of the rotating compact electrodynamics that we will discuss in the next section. 

\subsection{Rotation at imaginary angular frequency: \\
Wick rotation to imaginary time}

In this section, we consider the same system but at a purely imaginary angular frequency $\Omega_I$, Eq.~\eq{eq:Omega:I}. Our aim is to find a suitable representation of the scalar finite-temperature Green's function in the rotating background via analytical continuation from the imaginary angular frequency $\Omega_I$ to the real angular frequency $\Omega$. We will shortly see that the imaginary-frequency approach may also be suitable, besides applications in numerical lattice simulations, for certain analytical approaches as well.

We take the representation~\eq{eq:G:T:Omega} of the Green's function with imaginary frequency~\eq{eq:Omega:I}:
\beqn
& & G_{T,\Omega_I}({\bs x}, {\bs x}') = T \sum_{m,n \in \Z} \int\limits_0^{\infty} p d p \frac{e^{i \omega_n (\tau - \tau')}}{p^2 + (\omega_n + \Omega_I m)^2} 
\label{eq:G:T:Omega:I} \\
& & \times  \left[ \phi_{p,m}({\vec \rho}) \phi^*_{p,m}({\vec \rho}{\,}') - \phi_{p,m}\left(\frac{{\vec \rho} |{\vec \rho}{\,}'|}{R}\right) \phi^*_{p,m}\left(\frac{{\vec \rho}{\,}'_* |{\vec \rho}{\,}'|}{R}\right) \right],
\nonumber
\eeqn
and replace the sum over Matsubara frequencies $\omega_n$ by an integral over the continuous momentum $p_\tau$. To this end we substitute in Eq.~\eq{eq:G:T:Omega:I} the following identity:
\beqn
T \sum_{n \in \Z} f(\omega_n)= \sum_{l \in \Z} \int\limits_{- \infty}^{\infty} \frac{d p_\tau}{2\pi} e^{i p_\tau l/T} f(p_\tau),
\eeqn
and shift $p_\tau \to p_\tau - \Omega_I m$. We get:
\beqn
& & G_{T,\Omega_I}({\bs x}, {\bs x}') = \sum_{l,m \in \Z} \int\limits_{- \infty}^{\infty} \frac{d p_\tau}{2\pi} \int\limits_0^{\infty} p d p \frac{e^{i (p_\tau - \Omega_I m)(\tau - \tau')}}{p^2 + p_\tau^2} 
\nonumber\\
& & \times  \left[ \phi_{p,m}({\vec \rho}) \phi^*_{p,m}({\vec \rho}{\,}') - \phi_{p,m}\left(\frac{{\vec \rho} |{\vec \rho}{\,}'|}{R}\right) \phi^*_{p,m}\left(\frac{{\vec \rho}{\,}'_* |{\vec \rho}{\,}'|}{R}\right) \right],
\nonumber
\eeqn
The next step is to use the explicit form~\eq{eq:phi:pm} of the eigenfunctions $\phi_{p,m}({\vec \rho})$ as well as the plane wave expansion~\eq{eq:plane:wave} to show that
\beqn
& & \sum_{m \in \Z} \phi_{p,m}({\vec \rho}) \phi^*_{p,m}({\vec \rho}{\,}') e^{- i \Omega_I m(\tau - \tau' + l/T)}  = \int\limits_0^{2\pi} \frac{d \theta_p}{(2 \pi)^2}   \nonumber \\
& & \times e^{i p x \cos\left[ \theta_p + \theta  -\Omega_I \tau  \right] 
- i p x' \cos\left[ \theta_p + \theta' -\Omega_I (\tau' + l/T) \right] } \qquad \nonumber \\
& = & \int\limits_0^{2\pi} \frac{d \theta_p}{(2 \pi)^2}  e^{i {\vec p} \, [{\vec \rho}(\tau) - {\vec \rho}{\,}'(\tau' + l/T)]},
\label{eq:transformation}
\eeqn
where ${\vec \rho}(\tau)$ is defined in the following way:
\beqn
\rho_1(\tau) + i \rho_2(\tau) =  [\rho_1(0) + i \rho_2(0)]\,e^{- i \Omega_I \tau}.
\label{eq:vec:x:tau}
\eeqn
The interpretation of Eq.~\eq{eq:vec:x:tau} is straightforward: as the imaginary time $\tau$ increases, the spatial vector ${\vec \rho}(\tau)$ rotates about the origin  with the frequency equal to the imaginary angular frequency $\Omega_I$. 

The propagator reduces to
\beqn
G_{T,\Omega_I}({\bs x}, {\bs x}') & = & \sum_{l\in \Z} \int \frac{d^3 p}{(2\pi)^3} \frac{1}{{\bs p}^2} 
\label{eq:G:T:Omega:I:2}  \\
& &  \hskip -10mm \times \biggl\{  \exp \Bigl[ i {\bs p}  [{\bs x}(\tau) - {\bs x}'(\tau' + l/T)] \Bigr] \nonumber \\
& & \hskip -10mm - \exp \Bigl[ i {\bs p}  [{\tilde {\bs x}}(\tau) - {\tilde {\bs x}}'_*(\tau' + l/T)] \Bigr] \biggr\},
\nonumber\eeqn
where ${\bs p} = ({\vec p}, p_\tau)$ is the three-dimensional momentum,
\beqn
{\bs x}(\tau) = ({\vec \rho}(\tau),\tau),
\label{eq:x:tau}
\eeqn
is the spacetime coordinate with follows the helix~\eq{eq:vec:x:tau} in the imaginary time with the period $\Omega_I$, and
\beqn
{\tilde {\bs x}}(\tau) =\left( \frac{|{\vec \rho}{\,}'|}{R} {\vec \rho}(\tau),\tau\right).
\eeqn

The integral over the momentum $\bs p$ in Eq.~\eq{eq:G:T:Omega:I:2} can be easily taken~\eq{eq:D}. We get the following coordinate representation of the finite-temperature Green's function at purely imaginary rotational frequency $\Omega_I = - i \Omega$:
\beqn
& & G_{T,\Omega_I}({\bs x},{\bs x}') =  \frac{1}{4 \pi} \sum_{l \in \Z} \label{eq:D:Omega:I} \\
& & \times \biggl\{  \frac{1}{\bigl|{\bs x}(\tau) - {\bs x}'(\tau' + l/T)]\bigr|} - \frac{1}{|{\bs x}(\tau) - {\bs x}'(\tau' + l/T)]|_*} \biggr\} \nonumber \\
& &  =  \frac{1}{4 \pi} \sum_{l \in \Z} \Biggl[\frac{1}{\sqrt{ [{\vec \rho}(\tau) - {\vec \rho}{\,}'(\tau' + \frac{l}{T})]^{\,2} + \left(\tau - \tau '- \frac{l}{T}\right)^2}} \nonumber \\
& & \hskip 5mm - \frac{1}{\sqrt{  \frac{|{\vec \rho}{\,}'|^2}{R^2}  [{\vec \rho}(\tau) - {\vec \rho}{\,}'_*(\tau' + \frac{l}{T})]^{\,2}  + \left(\tau - \tau '- \frac{l}{T}\right)^2}} \Biggr],\nonumber
\eeqn
where ${\vec \rho}(\tau)$ is given in Eq.~\eq{eq:vec:x:tau} and the distance $|\dots|_*$ is defined in Eq.~\eq{eq:xy:2}. This Green's function vanishes if one of the arguments touches the boundary (if either $|\vec \rho| = R$ or $|\vec \rho{\,}'| = R$).

The interpretation of Eq.~\eq{eq:D:Omega:I} is straightforward: the imaginary frequency $\Omega_I$ makes the spatial coordinates $\vec \rho$ rotating about the origin with the angular frequency $\Omega_I$. The first term accounts for the direct propagation between the points while the second term takes into account a reflection at the cylindrical boundary. Thus, the real rotation in the real space is mapped to the imaginary rotation of the identical spacetime copies in the Wick-rotated Euclidean spacetime. A similar representation for the fermionic propagator can be found in Ref.~\cite{Ambrus:2019gkt}.

In the absence of the imaginary rotation, $\Omega_I = 0$, the helicoidal worldlines become straight lines and the Green's function~\eq{eq:D:Omega:I} expectedly reduces to Eq.~\eq{eq:D:0}.

Finally, we would like to stress that the Green's function~\eq{eq:D:Omega:I} is a periodic function of the imaginary time similarly to the Green's function in the absence of rotation~\eq{eq:G:periodicity}:
\beqn
G_{T,\Omega_I}\biggl({\bs x} {+} \frac{m}{T} \, {\bs{\mathrm{e}}_\tau}, {\bs y}\biggr) & = & 
G_{T,\Omega_I}\biggl({\bs x}, {\bs y} {+} \frac{m'}{T} \, {\bs{\mathrm{e}}_\tau}\biggr) \nonumber \\
& = & G_{T,\Omega_I}({\bs x}, {\bs y}),
\label{eq:G:Omega:periodicity}
\eeqn
where $m, m' \in \Z$. The periodicity~\eq{eq:G:Omega:periodicity} is obvious in the momentum representation~\eq{eq:G:T:Omega:I} and it may also be seen in the coordinate representation of the Green's function~\eq{eq:D:Omega:I}.

\subsection{High-temperature or long-distance limit}

\subsubsection{Green's function of a static disk}

In the high-temperature limit (or, equivalently, in the long-distance limit)
\beqn
T |{\vec \rho} - {\vec \rho}{\,}'| \gg 1,
\qquad
T |{\vec \rho} - {\vec \rho}{\,}'_*| \gg 1,
\label{eq:applicability:T}
\eeqn
the finite-temperature Green's function~\eq{eq:G:T:0} takes the following form (the leading term is shown only):
\beqn
G^{2d}_T({\vec \rho}, {\vec \rho}{\,}') 
= - \frac{T}{2\pi} 
\ln \left(  \frac{|{\vec \rho} - {\vec \rho}{\,}'|}{|{\vec \rho} - {\vec \rho}{\,}'_*|}\frac{R}{|{\vec \rho}{\,}'|}\right)\!,\ 
\qquad
\label{eq:G:T:2d}
\eeqn
which corresponds, up to the temperature factor $T$, to the Green's function of the two-dimensional Laplacian operator inside the disk of the radius $R$:
\beqn
\Delta^{(2d)}_{\vec \rho} G^{2d}_T({\vec \rho}, {\vec \rho}{\,}') = - T \delta({\vec \rho} - {\vec \rho}{\,}'),
\eeqn
where the mirror image point ${\vec \rho}_*$ in Eq.~\eq{eq:G:T:2d} is defined in Eq.~\eq{eq:mirror:point}. 

The Green's function~\eq{eq:G:T:2d} is expectedly symmetric with respect to the exchange of its arguments, $G^{2d}_T({\vec \rho}, {\vec \rho}{\,}') = G^{2d}_T({\vec \rho}{\,}', {\vec \rho})$. It also satisfies the Dirichlet conditions at the boundary: $G^{2d}_T({\vec \rho}, {\vec \rho}{\,}') = 0$ if either $|{\vec \rho}| = R$ or $|{\vec \rho}{\,}'| = R$. For a large disk, $R \to \infty$, the result~\eq{eq:G:T:2d} reduces ot the well-known expression in two dimensions~\eq{eq:log:0}.

\subsubsection{Green's function of a rotating disk}

Applying the same arguments to the Green's function of a disk~\eq{eq:D:Omega:I} rotating with the (imaginary) frequency~$\Omega_I$ we get:
\beqn
G^{2d}_{T,\Omega_I}({\vec \rho}, {\vec \rho}{\,}') & = & \frac{T}{4 \pi} \biggl[ H_{\Omega_I}\!\left(\rho, \rho', \theta  - \theta'\right) \nonumber \\
& & - H_{\Omega_I}\!\left(\frac{\rho \rho'}{R}, R, \theta  - \theta'\right) \biggr],
\label{eq:G:2d:inter:1}
\eeqn
where we use our standard notations $\rho_1 + i \rho_2 = \rho e^{i \theta}$, and defined the integral along the helix:
\beqn
H_{\Omega_I}(a, b, \theta) = \int\limits_{-\infty}^\infty \frac{d \xi}{\sqrt{\xi^2 {+} a^2 {+} b^2 {-} 2 a b \cos (\theta {+} \Omega_I \xi)}}. \qquad\
\label{eq:H:Omega}
\eeqn
Since the logarithmic divergence of this integral does not involve the physical parameters $a$, $b$, $\theta$ and $\Omega_I$, the divergence may easily be regularized.
In the absence of rotation, we recover Eq.~\eq{eq:G:T:2d} because
\beqn
H_0(a, b, \theta) = - \log (a^2 + b^2 - 2 a b \cos \theta) + C.
\eeqn

For the rotating system, the integral~\eq{eq:H:Omega} cannot be taken analytically. For small rotational frequencies $\Omega_I$, the integral can however be expanded in the series over $\Omega_I$ and the Green's function~\eq{eq:G:2d:inter:1} becomes as follows:
\beqn
G^{2d}_{T,\Omega_I} ({\vec \rho}, {\vec \rho}{\,}')  = G^{2d}_{T} ({\vec \rho}, {\vec \rho}{\,}') + G^{2d,(2)}_{T} ({\vec \rho}, {\vec \rho}{\,}') \Omega_I^2 + \dots, \qquad\
\eeqn
The first term in this expression is given in Eq.~\eq{eq:G:T:2d}.
To calculate the second term, we represent the integrand in Eq.~\eq{eq:H:Omega} via another integral,
\beqn
\frac{1}{\sqrt{A}} = \frac{1}{\pi} \int\limits_{-\infty}^{\infty} \frac{d \chi}{\chi^2 + \xi^2},
\eeqn
and change the integration over the cartesian coordinates $\chi$ and $\xi$ to the polar basis, $\chi + i \xi = \lambda e^{i \varphi}$:
\beqn
H_{\Omega_I}(a, b, \theta) & = & \int\limits_0^\infty d \lambda^2 \int\limits_{0}^{2 \pi} \frac{d \varphi}{2 \pi} 
\label{eq:H:Omega:2}\\
& & \times
\frac{1}{\lambda^2 {+} a^2 {+} b^2 {-} 2 a b \cos (\theta {+} \Omega_I \lambda \sin \varphi)}. 
\nonumber 
\eeqn
It is convenient to consider the difference of the $H$ functions in Eq.~\eq{eq:G:2d:inter:1} rather than the functions $H$ individually because this difference does not contain the logarithmic ultraviolet divergence. Then, expanding both terms~\eq{eq:H:Omega:2} in the imaginary frequency $\Omega_I$ and performing the straightforward integration, we obtain for the coefficient of the second term,
\beqn
G^{2d,(2)}_{T} ({\vec \rho}, {\vec \rho}{\,}') \equiv G^{2d,(2)}_{T} \left(\rho, \rho', \theta  - \theta'\right),
\eeqn
the following finite result (with $\theta_{\rho\rho'} \equiv \theta - \theta'$):
\beqn
& & G^{2d,(2)}_{T} (\rho, \rho', \theta_{\rho\rho'}) =  \frac{T}{4\pi} \left( \rho \rho' \!\cos\theta_{\rho\rho'} \right) 
\ln \left(  \frac{|{\vec \rho} - {\vec \rho}{\,}'|}{|{\vec \rho} - {\vec \rho}{\,}'_*|}\frac{R}{|{\vec \rho}{\,}'|}\right) \nonumber \\
& &  \quad + 2 \left(\rho^2 \rho'^2 \sin^2\theta_{\rho\rho'} \right) \left( 
\frac{1}{|{\vec \rho} - {\vec \rho}{\,}'|^2} - \frac{R^2}{|{\vec \rho} {-} {\vec \rho}{\,}'_*|^2 |{\vec \rho}{\,}'|^2} \right)\!. \qquad
\label{eq:G2:inter}
\eeqn

We wrote Eq.~\eq{eq:G2:inter} in the mixed notations:
\beqn
\rho \rho' \!\cos\theta_{\rho\rho'} & = & \frac{|{\vec \rho} - {\vec \rho}{\,}'|^2 - (\rho^2 + \rho'^2)}{2}, \\
\rho^2 \rho'^2 \!\sin^2\theta_{\rho\rho'} & = & \rho^2 \rho'^2  {-} \left(\frac{|{\vec \rho} - {\vec \rho}{\,}'|^2 {-} (\rho^2 + \rho'^2)}{2}\right)^2.\nonumber
\eeqn

Notice that the last line in Eq.~\eq{eq:G2:inter} is not divergent since for the close points with $\rho = \rho'$, one gets $|{\vec \rho} - {\vec \rho}{\,}'|^2 = 4 \rho^2 \sin^2 \theta/2 \sim \rho^2 \theta^2$ in the denominator of the first term, which is cancelled by the overall factor in the numerator, $\sin^2 \theta \sim \theta^2$. The last term in the second line is not divergent in the bulk. For growing separations between the points, $|{\vec \rho} - {\vec \rho}{\,}'|$, the second line in Eq.~\eq{eq:G2:inter} gives a subleading contribution with respect to the first line and thus the second line will be ignored in the following. 

Thus, the finite-temperature Green's function of the rotating disk at the imaginary momentum in a long-distance (or, equivalently, a high-temperature) limit takes the following form:
\beqn
G^{2d}_{T,\Omega_I}({\vec \rho}, {\vec \rho}{\,}') & = & - \frac{T}{2\pi} \left(1 - \frac{\Omega_I^2}{2} \rho \rho' \cos\theta_{\rho\rho'} \right)  \nonumber \\
& & \times \ln \left(  \frac{|{\vec \rho} - {\vec \rho}{\,}'|}{|{\vec \rho} - {\vec \rho}{\,}'_*|}\frac{R}{|{\vec \rho}{\,}'|}\right) + \dots,
\label{eq:G2d:Im}
\eeqn
where the ellipsis denote the omitted subleading terms.

Analytically continuing Eq.~\eq{eq:G2d:Im} to the real values of the orbital momentum, $\Omega_I^2 \to - \Omega^2$ we get the leading correction due to rotation to the Green's function in the large distance regime ($|{\vec \rho} - {\vec \rho}{\,}'| T \gg 1$):
\beqn
G^{2d}_{T,\Omega}({\vec \rho}, {\vec \rho}{\,}') & = & - \frac{T}{2\pi} \left(1 + \frac{\Omega^2}{2} \rho \rho' \cos(\theta - \theta') \right)  \nonumber \\
& & \times \ln \left(  \frac{|{\vec \rho} - {\vec \rho}{\,}'|}{|{\vec \rho} - {\vec \rho}{\,}'_*|}\frac{R}{|{\vec \rho}{\,}'|}\right) + \dots.
\label{eq:G2d:Re}
\eeqn
This formula will be used in the next section to analyze the effect of rotation on the confining properties of the system.

\section{Rotation and deconfinement}
\label{sec:confinement}

In this section, we consider the effect of the rotation on confinement and mass gap generation with the confining cU(1) gauge theory. Since any uniformly rotating system should be spatially bounded, we expect the appearance of the effects coming both from the presence of the boundary and from the effects of rotation. We discuss both these effects concentrating on the properties of the system in the bulk. 

\subsection{Deconfinement in a static system}

\subsubsection{Effect of boundaries on monopole density at $T=0$}

The presence of boundaries can modify the properties of the monopole gas even in the absence of rotation at zero temperature. Therefore, we expect that the boundaries may potentially affect the confining properties of the cU(1) gauge theory as well. In our work, we always consider systems which are much larger with respect to the Debye length~\eq{eq:m:ph} defined at a vanishing temperature,  $R \gg \lambda_D(T=0)$.

\begin{figure}[!thb]
\centerline{\includegraphics[scale=0.45, clip=true]{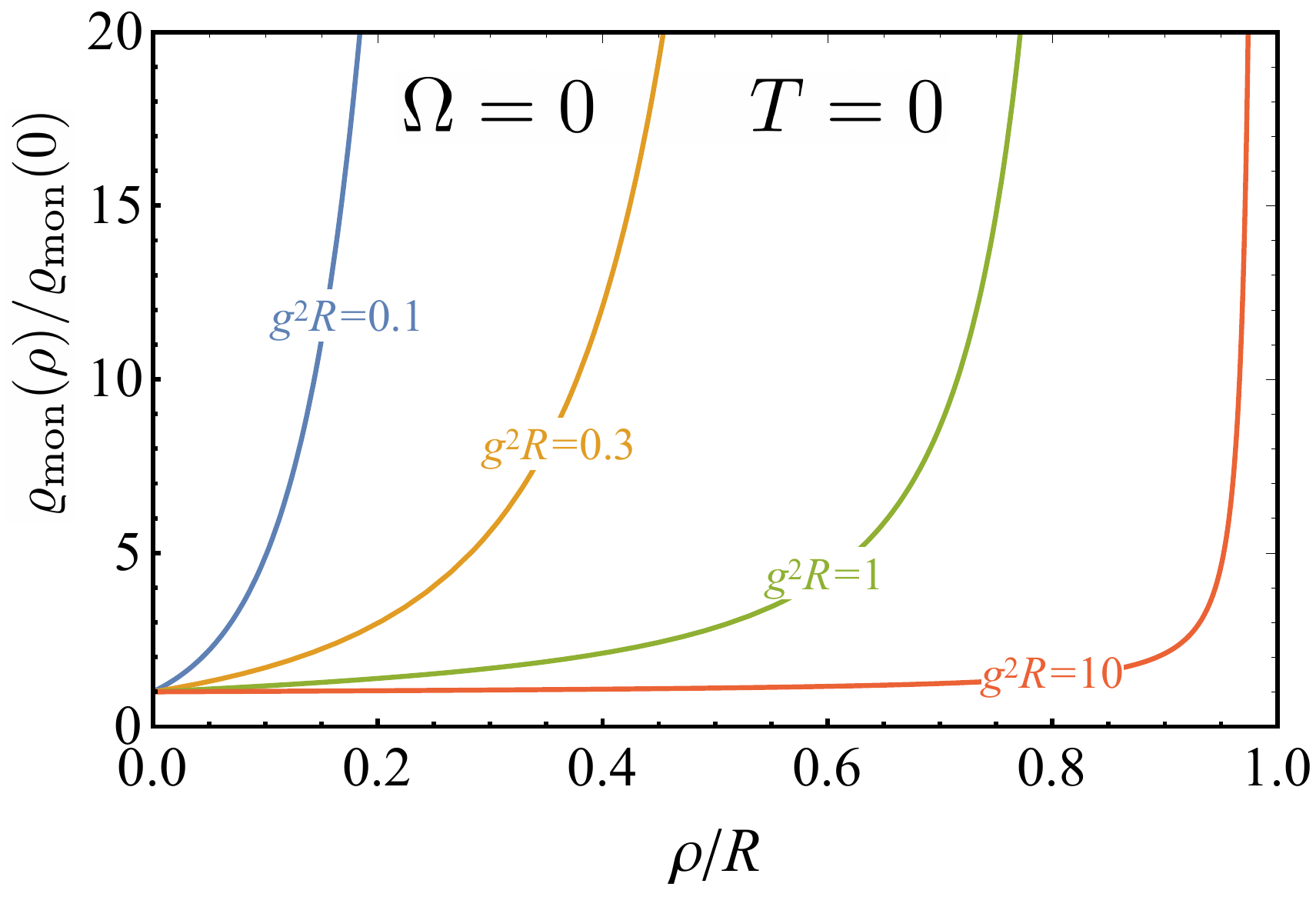}}
\caption{Monopole density~\eq{eq:rho:ratio} at zero temperature $(T=0)$ in the absence of rotation $(\Omega=0)$ as the function of the distance from the center $\rho$ at various values of the coupling constant $g^2$. The plots do not reflect correctly the properties of the density close to the boundary (with $R - \rho \ll R$) because in this region, the density diverges and the dilute gas approximation~\eq{eq:applicability} is no more valid.}
\label{fig:density:R}
\end{figure}

In order to estimate the effects of boundaries, temperature, and rotation on monopole density, we use the relation of the monopole density to the monopole fugacity in Eq.~\eq{eq:density}, the renormalization of the fugacity~\eq{eq:zeta:ren} via the Green's function, and the exact form of the Green's function~\eq{eq:G} at zero temperature. The ratio of the monopole density $\varrho_\mon(\rho)$ at the distance $\rho$ from the center of the disk to the monopole density computed at the very center of the disk is as follows:
\beqn
\frac{\varrho_\mon(\rho)}{\varrho_\mon(0)} = \exp \left( \frac{\pi}{2 g^2 } \frac{\rho }{R^2 - \rho^2} \right), \qquad (T = 0). \quad
\label{eq:rho:ratio}
\eeqn
The behavior of the monopole density for a set of coupling constants $g$ is shown in Fig.~\ref{fig:density:R}.

The density~\eq{eq:rho:ratio} becomes large as we approach the boundary, $\rho \to R$. This divergence is a natural property of the ``magnetic'' boundary characterized by the MIT condition~\eq{eq:MIT}: a positively charged monopole creates an image in the reflecting mirror and attracts to it, thus creating the excess of the monopoles (and anti-monopoles) close to the boundary. As one can see from Fig.~\eq{fig:density:R}, the stronger the coupling constant $g$, the stronger the attraction. The monopoles form pairs with their anti-monopole images (and vice-versa) at the boundaries.

The fact that the monopole density rises as we approach the boundary~\eq{eq:rho:ratio} implies that the string tension~\eq{eq:string:tension} and the Debye mass take their minima at the center of the disk and they both increase towards the border of the system:
\beqn
\frac{\sigma(\rho)}{\sigma(0)} = \frac{m_\ph(\rho)}{m_\ph(0)} = \exp \left( \frac{1}{4} \frac{\rho\, T_c}{R^2 - \rho^2} \right),\quad (T = 0). \qquad
\label{eq:sigma:ratio}
\eeqn
Our calculations are not valid in the very vicinity of the boundary because Eqs.~\eq{eq:rho:ratio} and \eq{eq:sigma:ratio} were obtained in the dilute gas approximation which is evidently broken at the edge of the system. In our calculations, we assume that we are not approaching the boundary too close, so that the applicability condition~\eq{eq:applicability} is still valid. We remind that the quantity $\varrho_\mon$ corresponds to the total number of the monopoles and antimonopoles regardless of their charges. The vacuum is magnetically neutral so that mean densities of monopoles and antimonopoles are equal to each other in every point.

\subsubsection{Boundary effects on monopole density at $T \neq 0$}

In the high-temperature limit (or long-distance limit) the Green's function~\eq{eq:G:T:0} is given by a combination of the logarithms~\eq{eq:G:T:2d}. Repeating the steps that led us to expression~\eq{eq:rho:ratio} but now with the dimensionally reduced Green's function~\eq{eq:G:T:2d}, we get the following behaviour of the monopole density as the function of the distance to the center $\rho$ of the disk:
\beqn
\frac{\varrho_\mon(\rho,T)}{\varrho_\mon(0)} = \left(\frac{R^2}{R^2 - \rho^2} \right)^\frac{T}{2 T_c},
\label{eq:rho:ratio:T}
\eeqn
where $T_c = g^2/\pi$ is the critical temperature of the deconfining BKT transition in the infinite volume~\eq{eq:Tc:inf}. The monopole density~\eq{eq:rho:ratio:T} increases towards the boundary, but at finite temperature this effect is much smoother as compared to the zero-temperature behaviour~\eq{eq:rho:ratio}. According to Eq.~\eq{eq:applicability:T}, the expression~\eq{eq:rho:ratio:T} is valid provided the distance between the point ${\vec \rho}$ and its image ${\vec \rho}_*$ is larger than the thermal length, $|{\vec \rho} - {\vec \rho}_*| T \ll 1$, or
\beqn
(R^2 - \rho^2) T \gg \rho.
\label{eq:applicability:T:R}
\eeqn

\begin{figure}[!thb]
\centerline{\includegraphics[scale=0.45, clip=true]{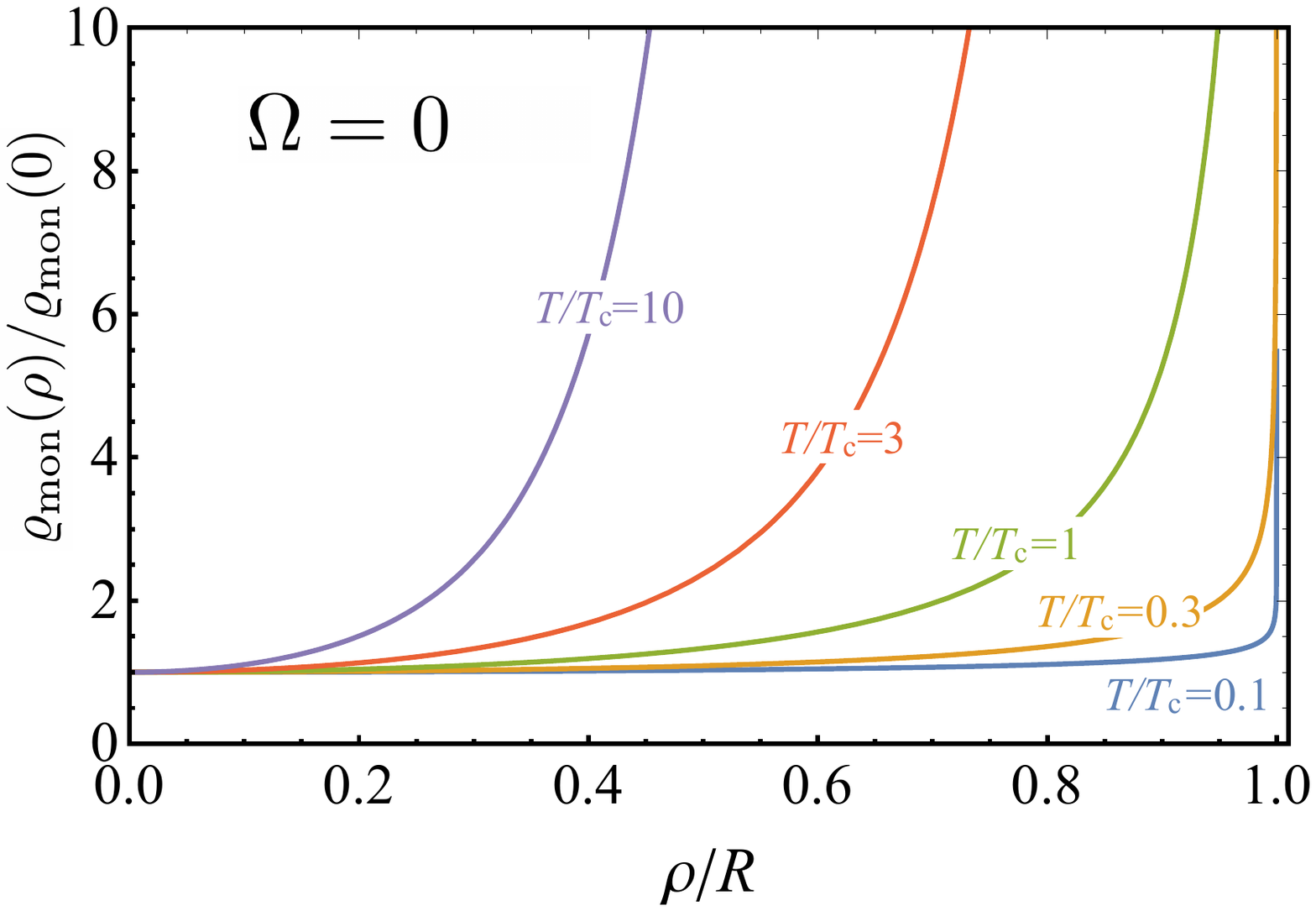}}
\caption{Monopole density~\eq{eq:rho:ratio:T} at various temperatures $T$ in a non-rotating plasma $(\Omega = 0)$. The applicability criterium~\eq{eq:applicability:T:R} implies that the behaviour close to the boundary cannot be trusted in the dilute gas approximation.}
\label{fig:density:T}
\end{figure}

The monopole density at zero temperature, Fig.~\ref{fig:density:R}, and the monopole density at finite temperature, Fig.~\ref{fig:density:T}, may be reconciled together by noticing that the weak-coupling limit, $g \to 0$, naturally corresponds to high values of the ratio $T/T_c$ at fixed temperature $T$. A similar statement is true for the opposite limit of the strong coupling. Notice, however, that due to the condition~\eq{eq:applicability:T:R}, a low-temperature limit $T \to 0$, the finite-temperature results can only be trusted sufficiently close to the rotation center, $\rho \to 0$.

\subsubsection{Effect of boundaries on deconfinement temperature}

We have already discussed the deconfinement transition in the static system in an infinite volume via the phase transition of the Berezinskii-Kosterlitz-Thouless (BKT) type with the critical temperature~\eq{eq:Tc:inf}.  At the critical temperature $T_c$, the instanton-like monopoles get bounded into magnetically neutral monopole-antimonopole pairs. At large distances, the magnetic field of the magnetically neutral monopole pairs falls down much faster compared to the field of individual monopoles, and the long-range confining property of the system is lost. 

In the disk-shaped space, the monopole action gets the form dictated by Eq.~\eq{eq:S:mon:T} and Eq.~\eq{eq:G:T:2d}:
\beqn
S^\mon = -  \frac{T g^2_\mon}{4\pi} \sum_{\stackrel{{a,b=1}}{a \neq b}}^N q_a q_b \, \ln \left(  \frac{|{\vec \rho}_a {-} {\vec \rho}_b|}{|{\vec \rho}_a {-} {\vec \rho}_{b,*}|}\frac{R}{|{\vec \rho}_b|}\right). \quad
\label{eq:S:Coulomb:gas:R:T}
\eeqn
The presence of the boundary does not modify the BKT transition temperature because the coefficient in front of the logarithmic term in the monopole action, $\ln|{\vec \rho} - {\vec \rho}{\,}'|$, is not modified. The nature of the phase transition should, however, be modified since the finite volume may not support a genuine phase transition. Usually, in a finite volume, phase transitions become smooth crossovers and the appropriate critical points are associated with the ``pseudo-critical'' rather than critical temperatures. We will continue to call them ``critical'' while keeping in mind their possible crossover nature.

The interaction potential gets the contribution due to the attractive interaction ($\ln |{\vec \rho} - {\vec \rho}{\,}'_*|$) of a monopole with its image~\eq{eq:G:T:2d} which gives a small polynomial correction to the energy of a monopole pair close to the center of the disk (with $|{\vec \rho}|, |{\vec \rho}{\,}'| \ll R$). Thus, we expect that the presence of the boundaries enhances the confinement closer to these boundaries without, however, modifying the deconfinement temperature of the BKT transition in the bulk.

In more detail, the numerator inside the logarithm of Eq.~\eq{eq:G:T:2d} corresponds to the attraction of monopoles and antimonopoles, and to the mutual repulsion of two monopoles and, separately, of two antimonopoles.  The denominator in Eq.~\eq{eq:G:T:2d} appears as the result of the presence of the boundary, as it describes the attraction of a monopole and an image of another monopole (and, likewise, with the anti-monopoles). The same denominator describes the repulsion of the anti-monopole and an image of the monopole, and vice versa. As we see from this expression, the denominator does not renormalize the temperature of the phase transition. 

One could therefore expect that the presence of the boundary diminishes the confining properties of the Coulomb plasma of monopoles and anti-monopoles since they appear together with their mirror images that have a screening effect on the magnetic field. In other words, the monopoles and anti-monopoles get self-screened by their images if they approach the boundary too close. Therefore, the confinement should become less pronounced closer to the boundary,  despite the fact that the monopole density~\eq{eq:rho:ratio:T} rises in the boundary's vicinity. However, this boundary-induced screening is indeed a purely boundary effect which should only be effective within one Debye distance~\eq{eq:m:ph} from the edge of the system, $(R - \rho) \sim \lambda_D$. We do not discuss it in the following, since we are interested in the bulk properties of the Coulomb plasma of monopoles.

We conclude that the bulk deconfinement temperature is not affected by the presence of the distant boundary (with $R \gg \lambda_D$) so that the critical temperature in the bulk of the disk coincides with the critical temperature in the infinite volume~\eq{eq:Tc:inf}.

\subsection{Deconfinement in a rotating system}

The rotation may, however, affect the phase structure of the theory. In this section, we address this question by studying the effect of the rotation on the monopole density and on the confining properties of the Coulomb gas of monopoles.

\subsubsection{Effect of rotation on monopole density}

Similarly to the case of the static system, the rotation affects the monopole density~\eq{eq:density} via the renormalization~\eq{eq:zeta:ren} of the fugacity parameter by the $\Omega$-dependent part of the Green's function~\eq{eq:G2d:Re}. Since the Green's function is known analytically only to the $O(\Omega^2)$ order, we use a numerical approach to evaluate the contribution of rotation to the Green's function~\eq{eq:cal:G} at large angular frequency $\Omega$. 

It is convenient to study the ratio of the densities,
\beqn
\frac{\varrho_\mon(\rho, T, \Omega)}{\varrho_\mon(\rho, 0, 0)} 
= \exp\left\{ - \frac{2 \pi^2}{g^2} {\mathcal G}_{\Omega,T}(\rho) \right\},
\label{eq:rho:ratio:Omega}
\eeqn
which quantifies the effect of the rotation on the monopole density.  

The quantity~\eq{eq:rho:ratio:Omega} evaluates the combined effect of the angular rotation and temperature at a fixed distance $\rho$ from the axis of rotation. This quantity has a different meaning as compared to the ratios studied previously: both the ratio~\eq{eq:rho:ratio} at zero temperature, shown in Fig.~\ref{fig:density:R}, and the ratio~\eq{eq:rho:ratio:T} at finite temperature, illustrated in Fig.~\ref{fig:density:T}, relate the densities at a finite distance $\rho$ with the density at the center of rotation. 

\begin{figure}[!thb]
\centerline{\includegraphics[scale=0.375, clip=true]{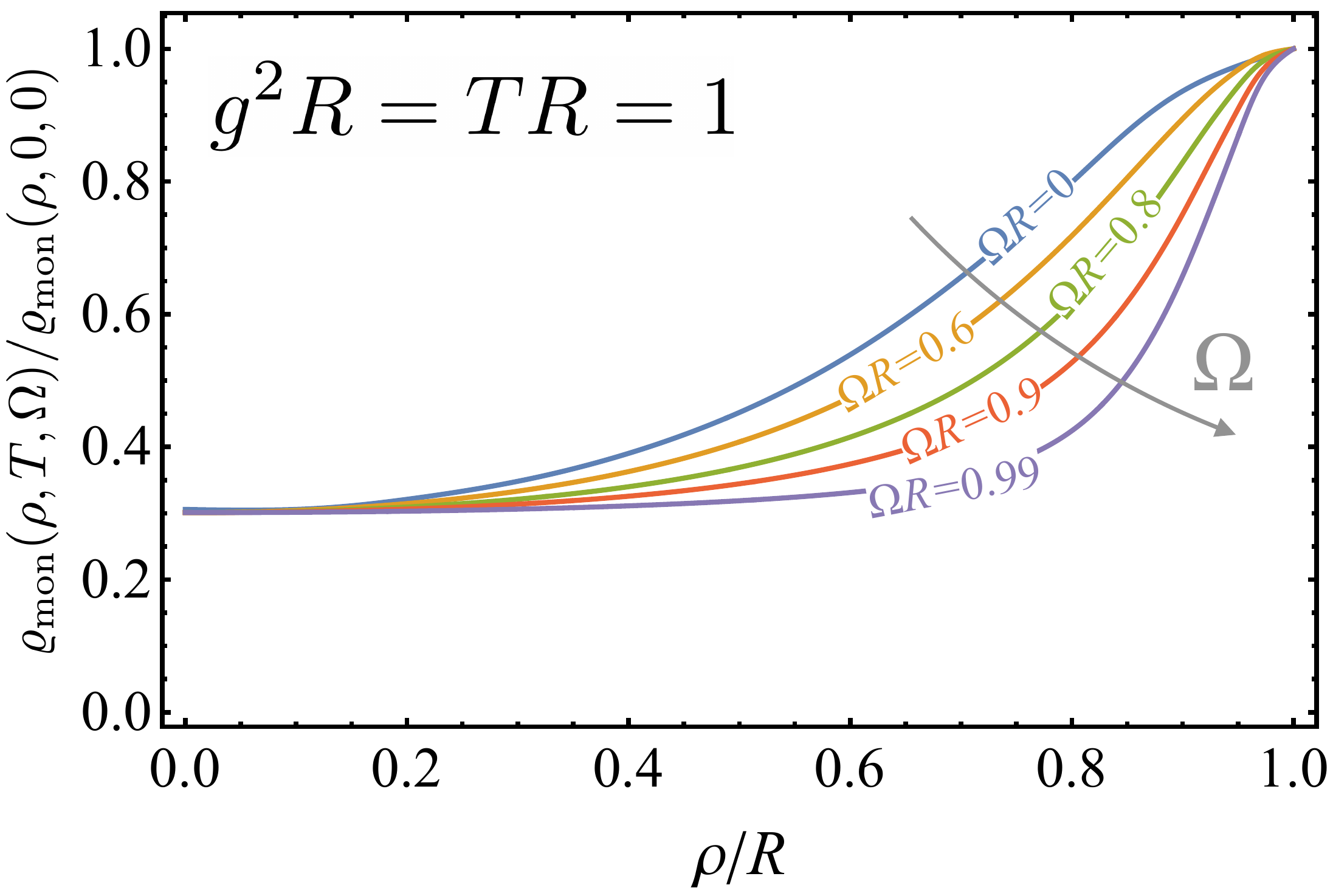}}
\caption{The ratio~\eq{eq:rho:ratio:Omega} of the monopole density at the angular velocity $\Omega$ and the temperature $T$ at the distance $\rho$ from the center of rotation to the monopole density at the same distance but for the vacuum at $T = \Omega = 0$.}
\label{fig:density:Omega}
\end{figure}

The behavior of the same-point propagator in the right-hand side of Eq.~\eq{eq:rho:ratio:Omega} has already been demonstrated in Fig.~\ref{fig:propagator}. The result of a calculation of the ratio~\eq{eq:rho:ratio:Omega} at fixed values of temperature $T$ and coupling $g$ is shown in Fig.~\ref{fig:density:Omega}. 

According to the definition of the Green's function~\eq{eq:cal:G}, the monopole density at the axis of rotation $\rho = 0$ does not dependent on the angular frequency $\Omega$ since $\varrho_\mon(0, T, \Omega) \equiv \varrho_\mon(0, T, 0)$. The density rises towards the boundary in agreement with the results of the previous section. As the angular velocity increases, the monopole density gets shifted towards the border of the system. The latter fact may be attributed to the centrifugal force acting on the monopoles. The effect of the angular frequency vanishes identically at the edge of the system due to the reflective nature of the boundary: the physical one-site Green's function~\eq{eq:cal:G} is zero at $\rho=R$.

\subsubsection{Effect of rotation on confinement}

As we have already mentioned, the value of the monopole density does not affect the critical temperature of the deconfinement phase transition. This conclusion, valid  in the dilute gas approximation, comes from the fact that the BKT mechanism bounds, at certain critical temperature, all monopoles and anti-monopoles into magnetically neutral pairs regardless of their initial density. The position of the critical transition is determined by the form of the interaction between the monopoles rather than by their quantity. 

The uniform rotation does indeed modify the prefactor in front of the logarithmic term of the Green's function~\eq{eq:G2d:Re} which enters the monopole action~\eq{eq:S:mon:T} and thus governs the finite-temperature interaction between the monopoles. Let us analyze this Green's function in detail, concentrating on the leading-order $O(\Omega^2)$ correction to the logarithmic interaction term.

First of all, we notice that the rotation does not influence the interaction between two (anti)monopoles if one of them is located exactly at the center of rotation ($\rho = 0$ or $\rho' = 0$). Mathematically, the effect appears because the rotational correction to the interaction~\eq{eq:G2d:Re} vanishes identically in this case. The effect of rotation is small for the monopoles which are located close to the rotation axis, with $\rho \Omega \ll 1$ or $\rho' \Omega \ll 1$. 

Far away from the center, the rotational effects becomes noticeable. Consider two nearby points ${\vec \rho}$ and ${\vec \rho}{\,}'$ such that $1 \ll |{\vec \rho} - {\vec \rho}{\,}'| T \ll \rho \Omega \simeq  \rho' \Omega$. In this case, we may set the angle $\theta$ between the vectors ${\vec \rho}$ and ${\vec \rho}{\,}'$ to zero in Eq.~\eq{eq:G2d:Re}, $\theta = 0$, an the Green's function takes the following form:
\beqn
G^{2d}_{T,\Omega}({\vec \rho}, {\vec \rho}{\,}') {=} {-} \frac{T_\Omega(\sqrt{\rho \rho'})}{2\pi} \ln \left(  \frac{|{\vec \rho} {-} {\vec \rho}{\,}'|}{|{\vec \rho} {-} {\vec \rho}{\,}'_*|}\frac{R}{|{\vec \rho}{\,}'|}\right) + \dots.\qquad
\label{eq:G2d:Im:2}
\eeqn
The effect of rotation may be incorporated in the spatial variation in the temperature, $T \to T_\Omega(\rho)$. The effective temperature,
\beqn
T_\Omega(\rho) = T(0) \left( 1 + \frac{1}{2} \rho^2 \Omega^2 + O(\Omega^4) \right)
\label{eq:T:eff}
\eeqn
acquires the dependence on the distance from the axis of rotation to the monopoles, $\rho \simeq \rho' \simeq \sqrt{\rho \rho'}$. The monopole action then becomes:
\beqn
S^\mon & = & -  \frac{g^2_\mon}{4\pi} \sum_{\stackrel{{a,b=1}}{a \neq b}}^N q_a q_b T_\Omega(\sqrt{\rho_a \rho_b}) 
\nonumber \\
& & \hskip 17mm \times \ln \left(  \frac{|{\vec \rho}_a - {\vec \rho}_b|}{|{\vec \rho}_a {-} {\vec \rho}_{b,*}|}\frac{R}{|{\vec \rho}_b|}\right).
\label{eq:S:Coulomb:gas:R:T:Omega}
\eeqn
In the absence of rotation, $\Omega = 0$, this action expectedly reduces to Eq.~\eq{eq:S:Coulomb:gas:R:T}.

What is the physical origin of the appearance of the effective temperature~\eq{eq:T:eff}? The temperature of a rotating physical body is defined in the co-rotating reference frame in which the body is static. In two spatial dimensions of the three-dimensional Minkowski spacetime, the co-rotating reference frame can be represented in terms of the curvilinear metric 
\begin{align}
g_{\mu \nu} =
\begin{pmatrix}
1-(x^2+y^2)\Omega ^2 & y\Omega & -x\Omega \\
y\Omega & -1 & 0 \\
-x\Omega & 0 & -1 \\
\end{pmatrix},
\label{eq:g:mu:nu}
\end{align}
with the line element (in the cylindrical coordinates):
\beqn
ds^2 & \equiv & g_{\mu\nu} dx^\mu dx^\nu \label{eq:metric} \\
& = & \left(1-\rho^2 \Omega^2 \right)dt^2- 2\rho^2\Omega dt d\varphi - d\rho^2- \rho^2 d\varphi^2 \,. \nonumber
\label{eq:ds2}
\eeqn
The rotation induces an effective gravitational field which is responsible, in particular, for the centrifugal forces. 

In a background gravitational field, the temperature  $T = T({\bs x})$  of a system in a thermal equilibrium is a local quantity defined by the Tolman-Ehrenfest law~\cite{Tolman:1930zza,Tolman:1930ona}:
\beqn
T({\bs x}) \sqrt{g_{00}({\bs x})} = T_0,
\label{eq:TE}
\eeqn
where $g_{00}$ is the component of the metric tensor. The reference temperature $T_0$ corresponds to a spatial point ${\bs x} = {\bs x}_0$ at which $g_{00}({\bs x}_0) = 1$.

For the rotating system~\eq{eq:g:mu:nu}, the relevant component of the metric tensor is $g_{00} = 1-\rho ^2 \Omega^2$. We find that the temperature~\eq{eq:T:eff}, which enters the monopole action, coincides exactly with the Tolman-Ehrenfest temperature~\eq{eq:TE} within the computed $O(\Omega^2)$ order. The reference quantity $T_0 \equiv T(0)$ is the local temperature at the axis of rotation, $\rho = 0$. Therefore, the influence of rotation reduces to the Tolman-Ehrenfest effect acting on the monopoles.

It is important to notice that it is the reference temperature $T_0$ -- and not the local temperature $T({\bs x})$ -- which is used in the imaginary-time formalism of the thermalized system and which is also implemented in the numerical simulations of lattice gauge theories~\cite{Yamamoto:2013zwa,Braguta:2020eis}.

The prefactor in front of the logarithmic term in the monopole action defines the phases of the monopole gas determining whether they are forming a Coulomb gas of individual objects (the confining phase) or are bounded into the neutral pairs (the deconfinement phase). Proceeding then along the lines that led us from Eq.~\eq{eq:S:mon:T} to Eq.~\eq{eq:Tc:inf}, we find that the coordinate-dependent prefactor of the monopole action~\eq{eq:G2d:Im:2} implies that in the rotating environment, the confinement is the coordinate-dependent property:
\beqs
\beqn
T_\Omega(\rho) & < & T_{c,\infty} \qquad \mbox{(confinement)}, \\
T_\Omega(\rho) & > & T_{c,\infty} \qquad \mbox{(deconfinement)}, 
\eeqn
\label{eq:T:rho}
\eeqs
where $T_{c,\infty} = g^2/\pi$ is the critical temperature in the thermodynamically large, nonrotating system~\eq{eq:Tc:inf}. The relations~\eq{eq:T:rho} should be understood in a quasi-local sense because the width of the spatial transition region between the two phases is of the order of the Debye length~\eq{eq:m:ph}.

Using the Tolman-Ehrenfest law~\eq{eq:TE}, supported by our calculations~\eq{eq:T:eff}, we find the critical line of the deconfinement transition in the temperature-radius plane:
\beqn
\left( \frac{T_c(\rho) }{T_{c,\infty}} \right)^2 + \Omega^2 \rho^2 = 1.
\label{eq:T:rho:2}
\eeqn
Since any physical system preserves the causality property, $\Omega^2 \rho^2 <1$, the critical temperature~\eq{eq:T:rho:2} is a well-defined quantity. 

Thus, we find that the uniformly rotating system possesses two transition temperatures:
\beqn
T_{c1} = T_{c,\infty} \sqrt{1 - \Omega^2 R^2}, \qquad\ T_{c2} = T_{c,\infty}. \qquad
\label{eq:Ts} 
\eeqn
The pure confinement phase is realized at $T< T_{c1}$. The mixed phase, which supports the confinement phase close to the rotational axis and the deconfinement phase in the outer layer, exists in between the upper and lower critical temperatures  $T_{c1} < T < T_{c2}$. The deconfining transition in the mixed phase appears at the following critical radius:
\beqn
R_c = \frac{1}{\Omega} \left( 1 - \frac{T^2}{T^2_{c,\infty}} \right)^{1/2} \,, 
\qquad T_{c1} < T < T_{c2}, \qquad
\eeqn
where we set $\Omega > 0$. Finally, the pure confining phase is realized at $T > T_{c2}$. According to Eq.~\eq{eq:T:rho:2}, the mixed phase disappears in the absence of rotation, because $T_{c1}(\Omega = 0) \equiv T_{c2}$.

From Eq.~\eq{eq:Ts} one can make an important conclusion: in a uniformly rotating system, the deconfining transition to the mixed phase may appear at any arbitrarily low, but still nonzero, temperature, provided the system rotates with a sufficiently large frequency, $\Omega R \sim 1$ (which does not exceed, however, the causality threshold $\Omega R < 1$).

\begin{figure}[!thb]
\centerline{\includegraphics[scale=0.45,clip=true]{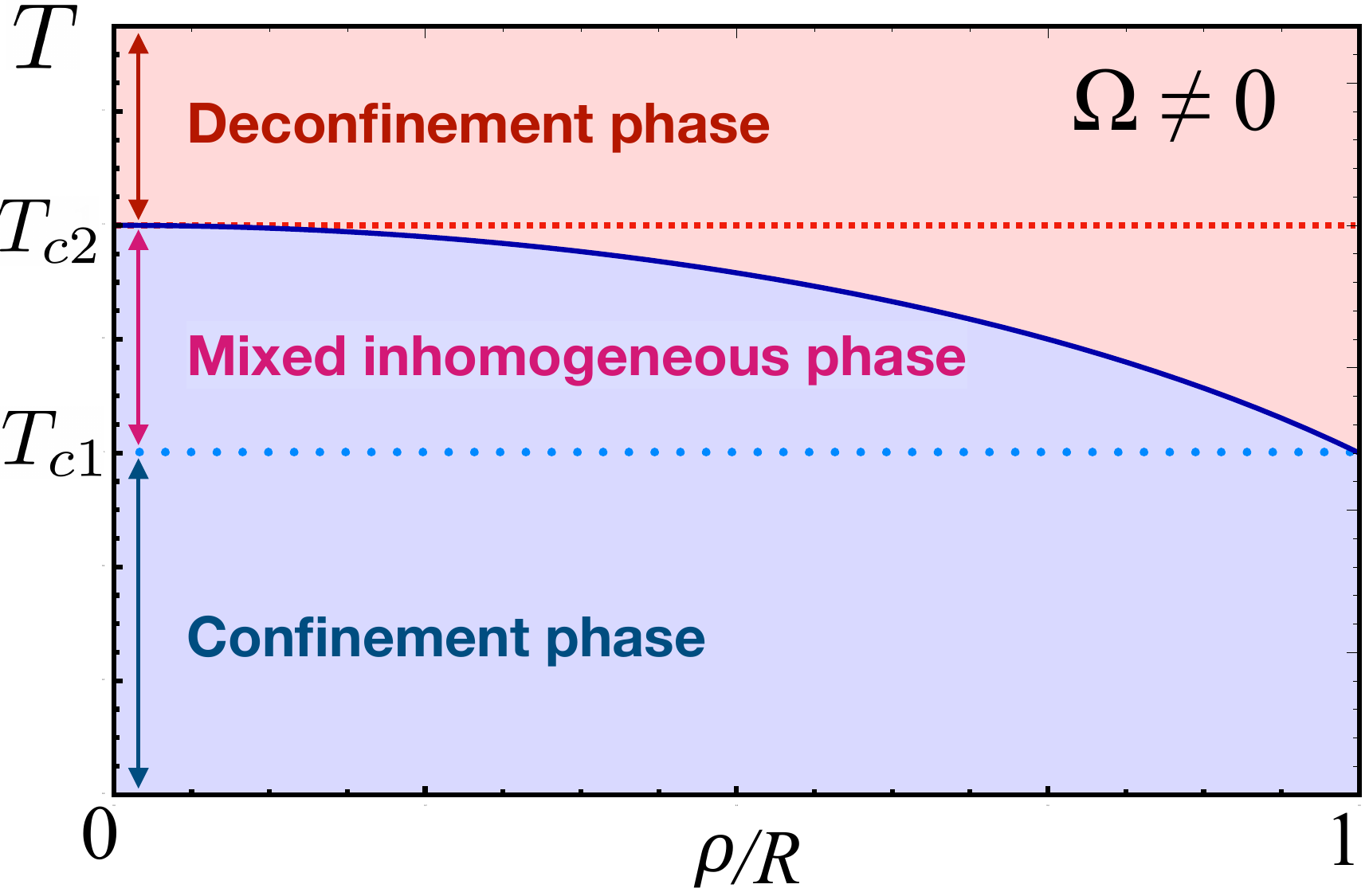}}
\caption{The local phase structure of the uniformly rotating confining field theory as a function of temperature $T$ and radius $\rho$ in a cylinder of a finite radius $R$ (with $R < 1/\Omega$).}
\label{fig:phase}
\end{figure}

\begin{figure}[!thb]
\centerline{\includegraphics[scale=0.25,clip=true]{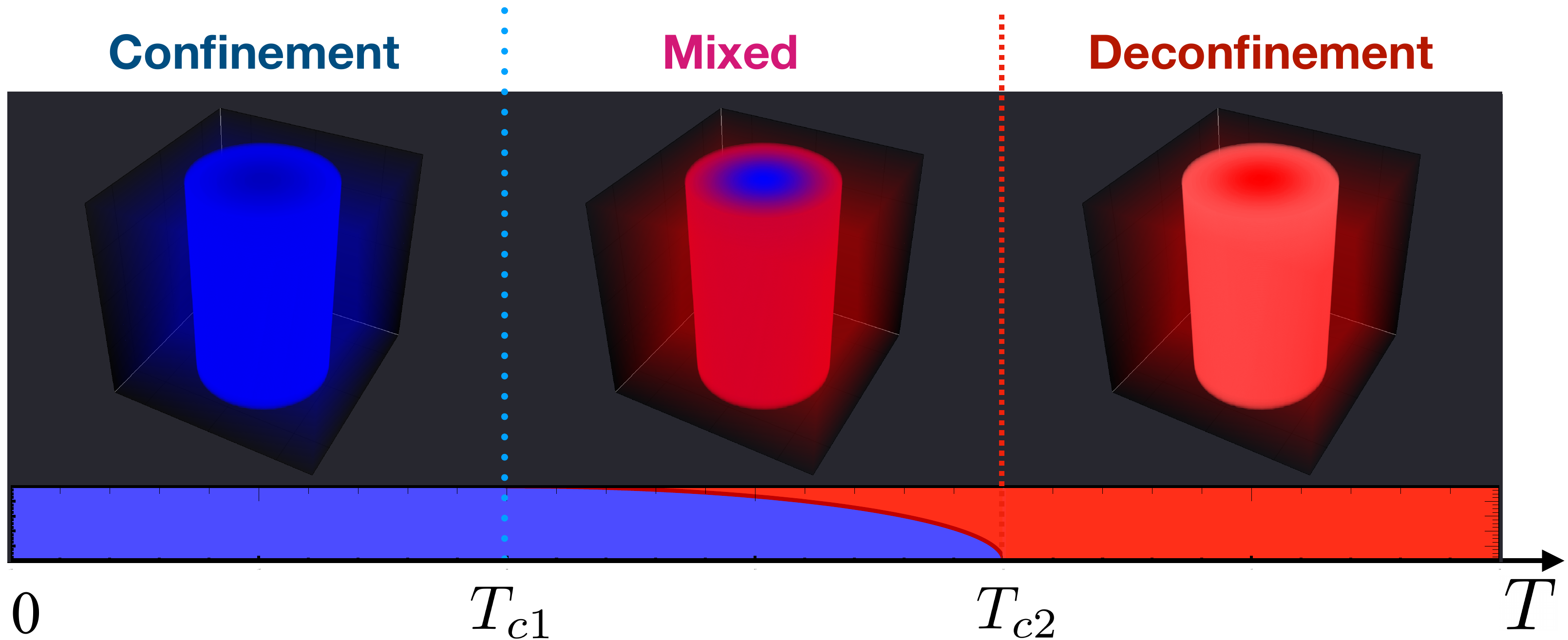}}
\caption{Illustration of the confining, mixed, and deconfining phases of the uniformly rotating system at finite temperature.}
\label{fig:phaseline}
\end{figure}

The phase diagram of the rotating confining system is shown in Fig.~\ref{fig:phase}, where the confinement, mixed, and deconfinement phases are presented. The spatial structure of these phases is illustrated in Fig.~\ref{fig:phaseline}. 

In a loose sense, the properties of (anti)monopoles under rotation may be interpreted as a result of the centrifugal force acting on these objects. The centrifugal force affects not only individual (anti-)monopoles via the increase of their density at the edges of the system, this force also modifies the interactions between the monopoles thus lowering the temperature of the deconfining transition. The deconfining effect increases as we approach the edge of the system. However, there is no effect of rotation neither on the deconfinement temperature nor on the density of monopoles at the very center of the rotating system. 

We expect that the same conclusions can also be applied to the theories in three spatial dimensions. It is difficult to speculate about confinement phenomenon in Yang-Mills theory starting from the first principles, but one could address the problem in the language of the effective theories which describe the confinement phenomenon. The rise of the kinetic temperature towards the edges of the rotating system will lead to the emergence of the mixed confining-deconfining phase, Fig.~\ref{fig:phaseline}, sliced in between the pure confining and pure deconfining phases of Yang-Mills theory.

\subsubsection{Comparison with existing results}

The only available first-principle lattice calculation of the deconfinement transition in pure SU(3) gluodynamics indicates that the rotation at the imaginary angular frequency $\Omega = i \Omega_I$ increases the expectation value of the bulk Polyakov loop at any fixed temperature around the deconfining phase transition~\cite{Braguta:2020eis}. The temperature, corresponding to the position of the susceptibility peak, decreases quadratically with the imaginary frequency $\Omega_I$. The latter property implies, via analytical continuation, that the critical temperature increases quadratically with the physical angular frequency~$\Omega$~\cite{Braguta:2020eis}. The results of the lattice simulations and the results in this paper are in contradiction with each other. 

The reasons for the discrepancy could be rooted in the property that the bulk Polyakov loop, calculated in Ref.~\cite{Braguta:2020eis}, acquires contributions from all the regions of the rotating plasma, including the central core (which a kinetically colder region) and the edges (which should have a higher kinetic temperature). Moreover, the Polyakov loop gets perturbative contributions which can depend on the rotation frequency. In contrast, we studied the local confining properties which were accessed not via the Polyakov loop but rather via the nonperturbative monopole properties. We have, however, a common qualitative (but not quantitative) agreement with the result of Ref.~\cite{Braguta:2020eis}: the critical transition~\eq{eq:Ts} depends on the angular momentum in the particular combination $\Omega R$, where $R$ is the radius of the system. 

Our results partially agree with the holographic estimation of the general deconfinement picture obtained in Ref.~\cite{Chen:2020ath}: the confining properties diminish with the increase of the angular velocity. However, in contrast with Ref.~\cite{Chen:2020ath}, we have found two distinct deconfining transitions and not a single one: the rotation splits the deconfining transition into two different deconfining transitions. Moreover, in our calculation, the first deconfining transition, from the confinement phase to the mixed phase, may be made arbitrarily low depending on the rotation frequency and on the size of the system while the second deconfining temperature does not depend on the rotation frequency at all (we expect that beyond the dilute gas approximation, the second deconfining temperature may get a weak dependence on the angular frequency).

\subsection{Phase diagram of rotating QCD}

Let us consider the effect of a pure kinematic rotation on the finite-temperature phase diagram of QCD at finite baryonic density. A uniformly rotating plasma can be described by the co-rotating curvilinear metric with the (3+1) dimensional line element:
\beqn
ds^2 & \equiv & g_{\mu\nu} dx^\mu dx^\nu = \left(1-\rho^2 \Omega^2 \right)dt^2  \label{eq:metric:3d} \\
& & - 2\rho^2\Omega dt d\varphi - d\rho^2- \rho^2 d\varphi^2 + d^2 z \,. \nonumber
\eeqn
The local temperature $T({\bs x})$ and the local chemical potential $\mu_B({\bs x})$ of the plasma subjected to the background gravitational field satisfy the following conditions~\cite{ref:LL:5}:
\beqn
T({\bs x}) \sqrt{g_{00}({\bs x})} = T_0,
\qquad
\mu_B({\bs x}) \sqrt{g_{00}({\bs x})} = \mu_{B0}, \quad
\label{eq:T:muB:x}
\eeqn
where $T_0$ and $\mu_{B0}$ are, respectively, the local temperature and the local baryonic chemical potential at the center of the rotating medium. The first relation in Eq.~\eq{eq:T:muB:x} is the Tolman-Ehrenfest law for temperature~\cite{Tolman:1930zza,Tolman:1930ona}, Eq.~\eq{eq:TE}.

According to Eq.~\eq{eq:T:muB:x}, the form of the zeroth component of the metric, 
\beqn
g_{00}(\Omega, \rho) = 1 - \Omega^2 \rho^2,
\label{eq:g00}
\eeqn
in the line element~\eq{eq:metric:3d} indicates that both the temperature and the baryon chemical potential rise as one moves from the center of the uniformly rotating plasma to its boundary. Therefore, the rotation kinematically shifts both $T$ and $\mu_B$ towards higher values and moves the QCD plasma from the confining, chirally-broken phase in the center of the system to the deconfining, chirally-restored quark-gluon plasma phase at outer layers of the rotating QCD matter. The region in the center of the plasma is unaffected by the rotation.\footnote{We remind that in our paper, we consider a uniform rotation of geometrically bounded plasma which is accessible, in the low-baryonic density domain, to first-principle numerical simulations of lattice QCD. The rotation of the quark-gluon plasma created in heavy-ion collisions is not expected to be uniform~\cite{Deng:2016gyh,Jiang:2016woz,Huang:2020dtn}.}

\begin{figure}[!thb]
\centerline{\includegraphics[scale=0.45,clip=true]{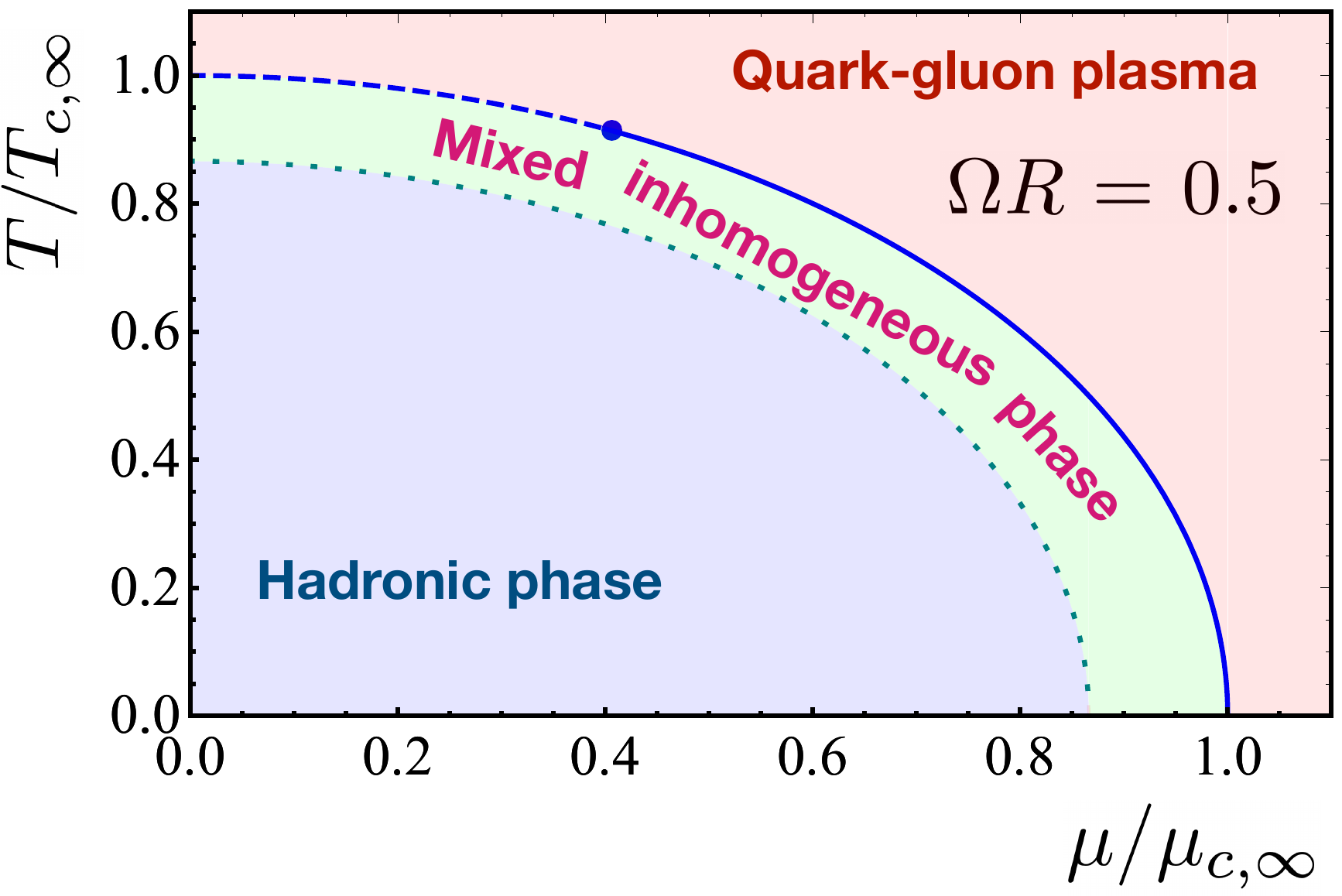}}
\caption{Suggested phase diagram of rotating QCD matter with $\Omega R = 0.5$ (for example, rotating with the angular frequency $\Omega = 0.1\,\mathrm{fm}^{-1}$ within the radius $R = 5\, \mathrm{fm}$). The chemical potential $\mu_B$ and the temperature $T$ are given at the geometrical center of the rotating plasma. The critical values $\mu_{c,\infty}$ and $T_{c,\infty}$ correspond to the thermodynamic limit of a non-rotating plasma. The position of the endpoint (the dot) which separates the 1st order phase transition (the solid line) and the crossover (the dashed line) is shown schematically. The first deconfinement transition is given by the dotted line. Higher-density and nuclear-matter transitions are not shown.}
\label{fig:phase:QCD}
\end{figure}

We stress that our discussion has a very general form based on the Tolman-Ehrenfest relations for temperature and chemical potential~\eq{eq:T:muB:x}. Therefore,  our conclusions are valid not only for the confinement phenomenon but they should also be applied to the chiral degrees of freedom as well.\footnote{Different boundary conditions can lead to both breaking~\cite{Zhang:2020hha} and restoration~\cite{Chernodub:2017ref} of the chiral symmetry at the very local vicinity of boundary. We ignore this skin effect because our discussion is devoted to the bulk properties.}

The kinematic effect of the increased temperature and the enhanced baryonic chemical potential~\eq{eq:T:muB:x} on the phase diagram of QCD can be figured out once we know the transition line $T_c = T_{c,\infty}(\mu_B)$ from the hadronic phase to the quark-gluon plasma phase in the thermodynamic limit of the nonrotating QCD matter. In Fig.~\ref{fig:phase:QCD} we show the QCD phase diagram under the uniform rotation. The diagram contains the mixed inhomogeneous phase, which possesses both the hadronic (confining and chirally broken) phase closer  to the center of rotation and the quark-gluon plasma (deconfining and chirally restored) phase in the rest of the volume. 

The line of the first finite-temperature transition from the pure hadronic phase to the inhomogeneous mixed phase is described as follows:
\beqn
T_{c1}(\mu_B,\Omega, R) = \sqrt{g_{00}(\Omega,R)} \, T_{c,\infty}\left( \frac{\mu_B}{\sqrt{g_{00}(\Omega,R)}} \right)\!, \qquad
\eeqn
where the metric element $g_{00}$ is given in Eq.~\eq{eq:g00}. This transition is shown by the dotted line without discriminating between the first order and crossover line. We expect that the finite-size environment will soften the strength of this transition and even make it a smooth crossover.

As the temperature and/or density rise, the island of the hadronic phase close to the center of rotation shrinks and finally disappears at the second (usual) transition line:
\beqn
T_{c2}(\mu_B,\Omega, R) = T_{c,\infty}(\mu_B).
\eeqn
We expect that this line is largely not affected by the kinematic rotation.

As the angular frequency increases, the inhomogeneous phase extends further into the confining phase in the phase diagram of Fig.~\ref{fig:phase:QCD}: the transition to the mixed phase emerges at lower temperatures and lower baryonic densities.

\section{Conclusions}

We found that a uniform rotation of a thermalized plasma of a confining gauge theory supports the appearance of a deconfinement transition in the regions far from the rotation axis. Therefore, at a finite temperature, a uniformly rotating system possesses three, rather than two phases: in addition to the confining phase at low temperature and deconfining phase at high temperature, the phase diagram contains also a mixed inhomogeneous phase at intermediate temperatures. The mixed phase has a confining region at the core and a deconfining region at the edge of the rotating system. The generic phase diagram is shown in Fig.~\ref{fig:phase} and illustrated in Fig.~\ref{fig:phaseline}. While we argue that this conclusion should have a universal character independent of the particularities of the confining theory, we supported our assertions by the explicit analytical calculations in compact Abelian gauge theory. In this model, the rotation directly affects the dynamics of Abelian monopoles and thus alters the system's confining properties.

The rotating plasma thus features two critical deconfining temperatures~\eq{eq:Ts} rather than the single one: the first critical temperature separates the confinement phase and the mixed phase while the second critical temperature marks the transition to the deconfinement phase.

Any uniformly rotating system should inevitably be bounded in the transverse (to the rotation axis) directions to support the causality. While the boundary conditions at the edge of the system could, in principle, affect the pseudocritical transition temperature(s), we have argued that the boundary conditions alone do not alter the transition temperature in the bulk (at least, in the dilute gas regime used in our article). 

We noticed that the rotation affects the monopoles in two different ways. First of all, the rotation has a centrifugal effect on the monopoles: it presses the monopoles against the rotating cylinder's edge. Secondly, the rotation enhances the attraction between monopoles and anti-monopoles, thus supporting the formation of the neutral monopole--anti-monopole pairs and facilitating the transition to the deconfinement phase in the region close to the boundary of the system. Thus, a uniformly rotating confining system possesses a mixed phase with both confinement and deconfinement phases separated by a spatial manifold (a ``deconfining cylinder'') where the deconfining transition is realized. Thus, instead of one, we have two transition temperatures which are separating confining, mixed, and deconfining phases of the rotating thermal medium. This behaviour is consistent with the Tolman-Ehrenfest law applied in the non-inertial frame which co-rotates with the system.

In a nonrotating system, the reflective MIT boundary condition, implemented for the gauge fields at the cylindrical boundary of the system, also leads to the increase of the monopole density close to the boundary and, therefore, to the enforcing the confinement properties in the low-temperature, confining phase. The explanation of this effect is simple: the monopoles and anti-monopoles are attracted to their images in the reflective mirror. The increased density of the monopoles enhances the confining properties close to the edge of the system which, however, does not affect the critical temperatures.

We have also shown that in the imaginary-time formalism, a uniform rotation is incorporated as a complex shift of the bosonic~\eq{eq:omega:b} and fermionic~\eq{eq:omega:f} Matsubara frequencies. The Wick-rotated angular frequency, $\Omega_I \equiv - i \Omega$ may be imagined as a ``rotation'' of the Euclidean system along the imaginary time which does not break, however, the (anti-) periodicity for (fermionic) bosonic fields along the imaginary time direction.

The partition function is invariant under an integer-multiple of a single $2\pi$--rotation for the bosons~\eq{eq:shift:b} and a double $4\pi$--rotation for fermions~\eq{eq:shift:f} in the agreement with the spin-statistics theorem. A bosonic Euclidean system which rotates with the imaginary frequency $\Omega_I = 2 \pi T$ (and with $\Omega_I = 4 \pi T$ for fermions) is equivalent to a stationary non-rotating field theory. These properties may be used to test the magnitude of the discretization effects in lattice implementations of the rotating field systems.

An unexpected outcome of this work is that a uniformly rotating, adiabatically expanding quark-gluon plasma should hadronize starting from its core and not from the outer layers as one could expect naively. This conclusion comes from the observation that the center of a uniformly rotating system is colder compared to its boundary. The effect of the ``inverse hadronization'' can be non-negligible for a real system: while the kinetic increase of temperature of the plasma rotating with the angular velocity $\Omega \sim 10\,\mbox{MeV}$ is expected to give a minuscule correction of $3\%$ at the radius $\rho \sim 5\, \mathrm{fm}$, the same effect leads to the very noticeable 30\% temperature rise for the $\Omega \sim 20\,\mbox{MeV}$ at the distance $\rho \sim 7\, \mathrm{fm}$ from the center. The appearance of the inhomogeneous mixed phase in the phase diagram of uniformly rotating QCD medium in the baryonic chemical potential $\mu_B$ -- temperature $T$ plane is illustrated in Fig.~\ref{fig:phase:QCD}. Since the rotation of the quark-gluon plasma is not globally uniform~\cite{Jiang:2016woz}, the effect of rotation on the confinement property in realistic physical environment requires a more detailed investigation.

\acknowledgments
The author is grateful to V.~E.~Ambru\c{s} and V.~V.~Braguta for interesting discussions. The work was partially supported by Grant No. 0657-2020-0015 of the Ministry of Science and Higher Education of Russia.

\end{document}